\documentclass[journal,10pt]{IEEEtran}

\usepackage{cite}
\usepackage{graphicx}
\usepackage{amsmath}
\usepackage{array}
\usepackage{mdwmath}
\usepackage{amssymb}
\usepackage{mdwtab}
\usepackage{stfloats}
\usepackage[tight,footnotesize]{subfigure}
\usepackage{amsmath,amsthm}
\usepackage{threeparttable}
\usepackage{color}
\usepackage{url}
\usepackage{algpseudocode}
\usepackage{algorithm}
\usepackage{multicol}
\usepackage{amssymb}
\usepackage{epstopdf}

\algrenewcommand{\algorithmicrequire}{\textbf{Input:}}
\algrenewcommand{\algorithmicensure}{\textbf{Output:}}

\newtheorem{theorem}{Theorem}
\newtheorem{remark}{Remark}
\newtheorem{lemma}{Lemma}

\ifCLASSINFOpdf
\else
\fi

\begin{document}

\title{Performance Analysis of RIS-Aided Double Spatial Scattering Modulation for mmWave MIMO Systems}

\author{
Xusheng Zhu, Wen Chen, \IEEEmembership{Senior Member, IEEE}, Qingqing Wu, \IEEEmembership{Senior Member, IEEE}, \\Jun Li, \IEEEmembership{Senior Member, IEEE}, Nan Cheng, \IEEEmembership{Member, IEEE}, Fangjiong Chen, \IEEEmembership{Member, IEEE},\\ and Changle Li, \IEEEmembership{Senior Member, IEEE}
\thanks{(\emph{Corresponding author: Wen Chen}).}
\thanks{X. Zhu, W. Chen, and Q. Wu are with the Department of Electronic Engineering, Shanghai Jiao Tong University, Shanghai 200240, China (e-mail: xushengzhu@sjtu.edu.cn; wenchen@sjtu.edu.cn; qingqingwu@sjtu.edu.cn).}
\thanks{J. Li is with the School of Electronic and Optical Engineering,
Nanjing University of Science Technology, Nanjing 210094, China (e-mail:
jun.li@njust.edu.cn).}
\thanks{N. Cheng and C. Li are with the School of Telecommunications Engineering, Xidian University, Xi'an 710071, China (e-mail: nancheng@xidian.edu.cn; clli@mail.xidian.edu.cn).}
\thanks{
F. Chen is with the School of Electronic and
Information Engineering, South China University of Technology, Guangzhou
510641, China (e-mail: eefjchen@scut.edu.cn).}
}

\markboth{}
{}

\maketitle
\vspace{-1.8cm}

\begin{abstract}
In this paper, we investigate a practical structure of reconfigurable intelligent surface (RIS)-based double spatial scattering modulation (DSSM) for millimeter-wave (mmWave) multiple-input multiple-output (MIMO) systems.
A suboptimal detector is proposed, in which the beam direction is first demodulated according to the received beam strength, and then the remaining information is demodulated by adopting the maximum likelihood algorithm.
Based on the proposed suboptimal detector, we derive the conditional pairwise error probability expression.
Further, the exact numerical integral and closed-form expressions of unconditional pairwise error probability (UPEP) are derived via two different approaches.
To provide more insights, we derive the upper bound and asymptotic expressions of UPEP. In addition, the diversity gain of RIS-DSSM scheme was also given.
Furthermore, the union upper bound of average bit error probability (ABEP) is obtained by combing the UPEP and the number of error bits.
Simulation results are provided to validate the derived upper bound and asymptotic expressions of ABEP.
We found an interesting phenomenon that the ABEP performance of the proposed system-based phase shift keying is better than that of the quadrature amplitude modulation.
Additionally, the performance advantage of ABEP is more significant with the increase of the number of RIS elements.
\end{abstract}

\section{Introduction}
In the past decade, numerous innovative wireless communication technologies have emerged to address the escalating demand for higher data rates and the rapid expansion of data services. For instance, massive multiple-input multiple-output (MIMO) systems enhance network throughput by serving multiple users simultaneously, and millimeter-wave (mmWave) communications utilize broader bandwidths to boost data rates over point-to-point links \cite{lu2014an,wang2017joi}.
Although these technologies have greatly improved the spectral efficiency of wireless communication, their inherent characteristics of high hardware cost and increased power consumption pose a competitive disadvantage in the development of next-generation wireless networks.

Recently, reconfigurable intelligent surface (RIS) has emerged as a promising technology for the sixth-generation (6G) wireless system, owing to its ability to create a customizable propagation environment at a low hardware cost and power consumption \cite{pan2021reconfigurable}. Essentially, the RIS comprises a two-dimensional meta-surface consisting of numerous low-cost reflecting elements, which can individually reflect incoming signals with adjustable phase shifts \cite{wu2019intelligent}.
It is worth noting that the phase shifts can be adjusted by employing inexpensive negative diodes or varactor diodes via the RIS controller, which can manipulate the electromagnetic waves, thus improving the signal quality of service and extending the network coverage \cite{li2022joint}.
Unlike conventional transmit/reflect-array antennas, the RIS can be conveniently deployed to facilitate data transmission without requiring any power-hungry transmit radio-frequency (RF) chains for signal reception and transmission, thus significantly cutting hardware costs and energy consumption \cite{wu2020intelligent}.
Explicitly, the RIS technology is initially advocated in single-input single-output (SISO) systems, where the passive RIS can be employed to reflect the incoming signal \cite{Basarwirl}.
Although the introduction of RIS can significantly reduce the average bit error rate (ABEP) of the system, the transmission throughput of the SISO-assisted RIS structure is limited in meeting the growing demand for high throughput in future wireless communications.
To enhance the transmission throughput, MIMO-based RIS architectures have been investigated {\cite{Yigit2020low}}. Specifically, the initial efforts focused on channel estimation and passive beamforming, where the RIS modifies the amplitude and phase of the incident signal. With an increased number of antennas, the transmission throughput of the MIMO-RIS technology can be greatly improved \cite{pan2020int}. However, the augmented cost and the growing complexity of signal processing exacerbate the practical implementation. Accordingly, further design of simplified MIMO-RIS structures is needed.

\begin{table*}[t]
\centering
\caption{\small{Notations in this paper}}
\begin{tabular}{|c|c|}
\hline Notations & Definitions \\
\hline$\|\cdot\|$ & Euclidean norm of a complex-value vector \\
\hline $|\cdot|$ & Absolute value operation \\
\hline$(\cdot)^H$ & Conjugate transpose operator \\
\hline $(\cdot)^T$ & Transpose operator \\
\hline $\mathcal{CN}(\mu,\sigma^2)$ & Circularly symmetric complex Gaussian (CSCG) distribution \\
\hline $\mathcal{N}(\mu,\sigma^2)$ & Real Gaussian distribution\\
\hline $\mathbb{C}^{n\times m}$ & The space of $n\times m$ compelx-valued matrics \\
\hline $\sim$ & ``Distributed as" \\
\hline $ {\rm diag}(\cdot)$ & Diagonal matrix operation  \\
\hline $\arg[\cdot]$ & Phase operations on complex numbers \\
\hline $P_s$ & The average transmit power\\
\hline $\delta(\cdot)$ & Delta function  \\
\hline $\exp\left(\cdot\right)$ &Exponential function\\
\hline$\Pr(\cdot)$ & Probability of the event occurring \\
\hline $|\cdot|$ & The absolute value operation \\
\hline $P_b$ & CPEP \\
\hline $\bar P_b$ & UPEP\\
\hline $\Re\{\cdot\}$ & Real part operation \\
\hline $\Im\{\cdot\}$ & Imaginary part operation \\
\hline $E(\cdot)$ & Expectation operation \\
\hline $\Gamma(\cdot)$ &  Gamma function \\
\hline $W_{\cdot,\cdot}(\cdot)$ & Whittaker function \\
\hline $K_{0}(\cdot)$ &   Zeroth order modified Bessel function of second kind\\
\hline $K_{1}(\cdot)$ & First order modified Bessel function of second kind \\
\hline $I_0(\cdot)$ & Zero-order modified Bessel function of the first kind \\
\hline $Q(\cdot)$& Q-function \\
\hline $\psi$  & Digamma function \\
\hline$\rm erfc(\cdot)$ & Complementary error function \\
\hline$(\cdot)!$ & Factorial operation \\
\hline
\end{tabular}
\end{table*}

Spatial modulation (SM) is a widely investigated technology, as evidenced by previous studies \cite{mes2008spati,zhu2022on,jeg2009spac,meslen2015qua}. The information conveyed by SM consists of two components, the spatial domain and the symbol domain, thereby enhancing spectral efficiency.
It is worth noting that SM enables a transmitter (Tx) to use only one radio frequency (RF) chain, which can be activated by selecting a specific transmit antenna for each time slot based on the input bits.
Since only one transmit antenna is active at each time slot and all others are inactive, SM strikes a trade-off between energy efficiency and spectral efficiency \cite{mes2008spati}.
To investigate the propagation characteristics of SM in the real-world environment, the authors of \cite{zhu2022on} performed large-scale measurements to extract the channel impulse responses in the lecture room and indoor corridor under the line-of-sight (LoS) and non-line-of-sight (NLoS) scenarios, respectively.
For analysis of the spatial domain information performance of SM, the authors of \cite{jeg2009spac} introduced the concept of space shift keying (SSK), in which the signal of the symbol domain is stripped out and relies only on the signal of the spatial domain to complete the transmission.
In particular, the authors of \cite{meslen2015qua} presented the quadrature SM (QSM) scheme to increase the spectral efficiency by extending the SM into in-phase and quadrature dimensions.
All of the above work is based on activating antennas at different locations in space to achieve information transmission, however, such techniques are hardly applicable in the mmWave band.
Since the high-frequency wavelength of mmWave is quite small, this results in higher path loss in the propagation of information.
Accordingly, it is difficult to efficiently deliver the desired signal to the target Rx through a single antenna in the mmWave band with SM and variants of SM techniques.
On the other hand, the scarcity of low-band spectral resources, mmWave has a much wider band range, which effectively alleviates the spectral resource constraint.
In view of this, spatial scattering modulation (SSM) scheme is proposed in \cite{ding2017ssm}, where the transmit beam of the SSM scheme can be applied at each transmission time slot instead of a single antenna, thus guaranteeing the quality of the signal.
In contrast to conventional SM schemes, the SSM also employs a single RF chain. However, instead of activating a transmit antenna at each time slot, the SSM activates a scatterer in the channel.
In \cite{li2019polaried}, the quadrature SSM scheme was studied, where the real and imaginary parts of the transmit signal are transmitted using different transmit beams, thereby enhancing spectral efficiency.

\begin{figure*}[t]
\centering
\includegraphics[width=16cm]{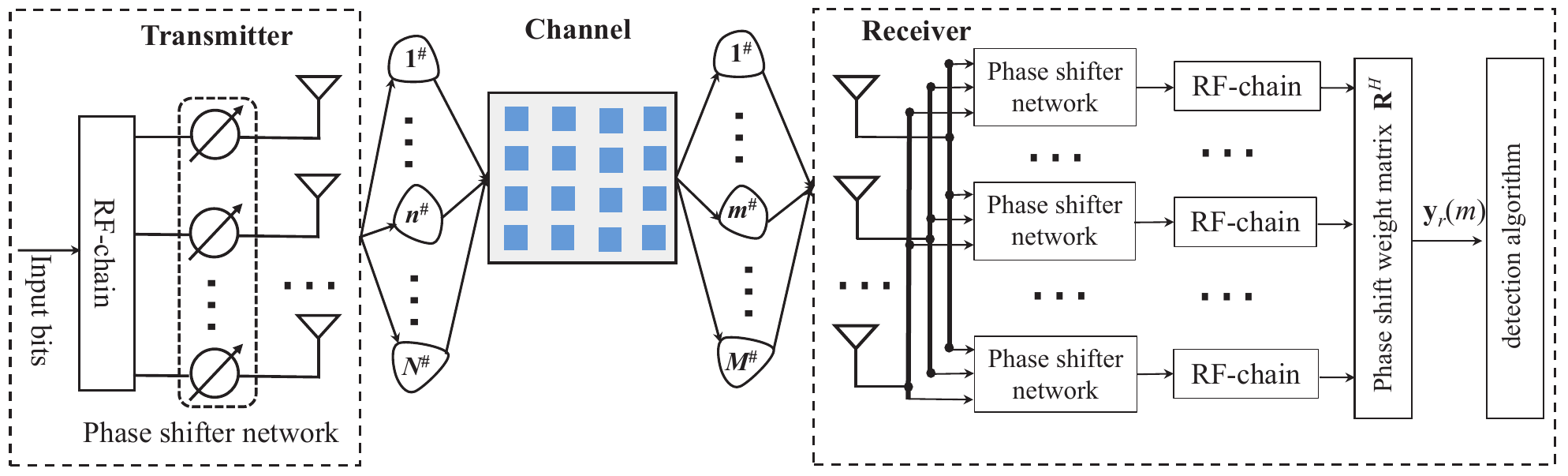}
\caption{\small{System model of the proposed RIS-DSSM scheme.}}
\vspace{-10pt}
\label{sys}
\end{figure*}

\subsection{Related Works}
Driven by the enhanced coverage of RIS and the improved spectral efficiency features of SM, the combination of the two has the potential to be an energy-efficient communication solution \cite{easar2020rec,ma2020large,yuan2021recei,bho2021ris,luo2021spatial}.
For instance, the author of \cite{easar2020rec} firstly modeled an RIS-assisted SM scheme, in which the RIS-SM scheme can acquire maximum signal-to-noise ratio (SNR) by adjusting the phase shifts of the RIS.
Different from \cite{easar2020rec} only focused on the receive SM, the authors of \cite{ma2020large} studied the large intelligent surface-assisted SM (LIS-SM) scheme by adopting SM both at the Tx and receiver (Rx).
Besides, optimization problems aimed at minimizing the symbol error rate were respectively investigated under the transmitting SM and receiving SM scenarios \cite{luo2021spatial}.
In \cite{yuan2021recei}, Wu \emph{et. al} concentrated on the modulation scheme for RIS-assisted symbiotic radio systems, where the primary and additional internet-of-things signals are jointly detected at the Rx.
Additionally, the authors of \cite{bho2021ris} studied an energy-efficient architecture that leverages SM technology and RIS-assisted ambient backscatter (ABSc) communication. This architecture employs ABSc to capture energy from ambient RF waves, reducing power consumption and extending the battery life of wireless devices.
Furthermore,
\cite{lin2022reconf} proposed a novel reflection modulation scheme named RIS-assisted quadrature reflection modulation (RIS-QRM), which is used for simultaneous passive beamforming and information transmission.
In particular, the works on RIS-assited SSK schemes were investigated in \cite{can2020rec,din2022ris,li2021space,zhu2023per,zhu2023ris,zhu2023imsk}.
For example, the performance of ABEP on RIS-SSK schemes was studied in \cite{can2020rec,zhu2023per,zhu2023imsk}. To further improve the spectral efficiency, the authors of \cite{li2021space} derived the closed-form expression of ABEP and presented an RIS-SSK scheme with Alamouti space-time block coding.
Unlike \cite{li2021space}, the authors of \cite{din2022ris} investigated the RIS-assisted receive quadrature space shift keying (RIS-RQSSK) scheme that increases the spectral efficiency by employing the real and imaginary dimensions independently.
To improve the spectral efficiency of the RIS-SSK system, the RIS-assisted full-duplex SSK scheme was studied in \cite{zhu2023ris,zhu2023p1}.

\subsection{Motivation and Contribution}
In this paper, we consider a point-to-point RIS-aided modulation for mmWave systems with large-scale antenna arrays at the Tx and Rx sides.
Due to unfavorable channel conditions, the direct link between Tx and Rx is blocked out. For this reason, we resort to RIS to assist the communication between Tx and Rx.
To fully exploit the gain of RIS for modulation systems in the mmWave band, it is essential to investigate the RIS-assisted SSM system, which has potential applications in secure communication. Concretely, the code represented by the participating scatterers is known only to Tx and Rx. Meanwhile, it is very hard for an eavesdropper to recognize the code by the participating scatterers, especially in mobile communication scenarios where the scatterers are continuously changing.

Up to now, research on RIS-assisted SSM transmission has been still in its infancy.
Fortunately, there are similar studies such as the work presented in  \cite{zhu2023mmwave,zhu2023on,zhu2022recon}.
To be specific, \cite{zhu2023mmwave} and \cite{zhu2023on} considered a scenario in which the RIS is installed closer to the Tx and farther away from the Rx, where the Tx-RIS and RIS-Rx channels exchange information via line-of-sight (LoS) and non-LoS (NLoS) paths, respectively.
Moreover, \cite{zhu2022recon} investigated a more general RIS-double SSM (SSM) model, in which both Tx-RIS and RIS-Rx channels are communicated with the NLoS paths.
In particular, \cite{zhu2022recon} considered that the scatterer gains of the two sub-channels are arranged in descending order, which significantly increases the complexity of deriving the probability density function (PDF) of the cascade channel, rendering the problem too complicated to address.
With respect to performance analysis, \cite{zhu2022recon} only provides conditional pairwise error probability (CPEP) expressions without further derivation of unconditional pairwise error probability (UPEP) expressions.
Furthermore, due to the high complexity of maximum likelihood (ML) detectors, \cite{zhu2022recon} does not provide a low-complexity detection algorithm.
To this end, this paper proposes a novel detection and decoding algorithm to comprehensively evaluate the performance of RIS-assisted DSSM systems.
The main contributions of this work can be summarized as follows:
\begin{itemize}
\item We provide a new analytical framework on the performance analysis of RIS-DSSM scheme.
    A suboptimal detection algorithm is proposed in this paper and we compare it with the ML detection algorithm in terms of complexity and reliability.
\item Considering a practical model for the phase shift and amplitude response, we derive exact integral expressions for the ABEP. To get additional insights into the impact of system parameters, we derive the closed-form and asymptotic expressions for the ABEP.
    Additionally, the diversity gain of the proposed RIS-DSSM scheme is also provided.
\item Simulation results are in agreement with the analytical ABEP based on the exact integral expression. Besides, we adopt the derived integral expression to validate the asymptotic expression for the RIS-DSSM scheme.
    Furthermore, the RIS-DSSM scheme has better ABEP performance compared to the conventional SSM scheme when the number of RIS elements is relatively large.
\end{itemize}

The remaining sections of this paper are structured as follows. In Section II, we introduce the system model and propose a low-complexity suboptimal detector for the RIS-DSSM scheme. In Section III, the exact integral and closed-form expressions of UPEP are derived. Based on this, the union upper bound of ABEP is provided.
After that, Simulation and analytical  results are presented and discussed in Section IV. Finally, Section V gives a summary of this paper.
Note that the notations of this paper are summarized in Table I.

\section{System Model}
In this paper, we consider a point-to-point mmWave MIMO system shown in Fig. \ref{sys},
where the Rx and Tx are equipped with $N_r$ and $N_t$ element uniform linear arrays (ULAs), respectively. Meanwhile, the RIS is deployed to assist the data transmission from the Tx to the Rx, and RIS is composed of $L$ reflecting elements.
To characterize the limited scattering feature, we adopt the Saleh-Valenzuela (S-V) model to denote the sparse scattering of mmWave wireless propagation.
There is a specific beam pointed to the candidate scatterer by phase shifter at each time slot.
Since a single RF link is equipped at the Tx, only one transmitted beam can be generated to align the target scatterer.
In other words, the transmit beam can only be aimed at the scatterers in the Tx-RIS channel or the scatterers in the Tx-Rx channel in each time slot.
If the beam target the scatterer in the Tx-Rx channel, it becomes the conventional SSM schemes, which has been  studied in \cite{ding2017ssm}.
In particular, if the RIS is far from the Tx or Rx, the effect of RIS assistance can be ignored. But this is not the consideration of this paper, since our goal is to investigate the performance of how much RIS can help SSM, where RIS is deployed not too far away from the Tx or Rx.

In this paper, we assume that Tx, Rx, and RIS can obtain perfect channel state information (CSI).
However, in practical situations, channel estimation is unavoidable. Considering this, the RIS-DSSM scheme under ideal channel conditions can be taken as a limit for the actual channel estimation. When the channel estimation error is considered, the performance of the RIS-DSSM system inevitably deteriorates, and this part of the study can be left to the interested reader to expand on this foundation.

\subsection{RIS-DSSM Transmission}

In the RIS-DSSM scheme, the information stream consists of three parts, i.e., spatial beam stream in the Tx-RIS channel, spatial beam stream in the RIS-Rx channel, and symbol stream.
Since the Tx employs a single RF chain, the beam is directed to a specific scatterer among the $N$ scatterers at each time slot in the Tx-RIS channel. For the spatial domain signal stream in the RIS-Rx channel, the RIS controller can obtain perfect CSI of the Tx feedback so that the RIS can reflect the incident signal and be steered to a candidate scatterer among the $M$ scatterers.
Furthermore, the symbol domain signal adopts the $K$-ary phase shift keying/quadrature amplitude modulation (PSK/QAM).
As a result, the total data rate for the proposed RIS-DSSM transmission can be represented as \cite{li2019polaried}
\begin{equation}
R = \log_2(N) + \log_2(M) + \log_2(K).
\end{equation}
\subsection{RIS-DSSM Channel Model}
In Fig. \ref{sys}, the RIS divides the channel into Tx-RIS channel and RIS-Rx channel.
The channel maritx $\mathbf{H}$ can be evaluated as
\begin{equation}\label{hh1}
\mathbf{H} = \mathbf{H}_{\rm TI} \boldsymbol{\Phi} \mathbf{H}_{\rm IR},
\end{equation}
where the phase-shifter matrix $\boldsymbol{\Phi}$ at RIS can be expressed as
\begin{equation}\label{hh2}
\boldsymbol{\Phi} = {\rm diag}(e^{j\phi_1},\ldots,e^{j\phi_L}) \in \mathbb{C}^{L \times L},
\end{equation}
where $\phi_l \in [0, 2\pi]$ represents the phase shift caused by the $l$-th RIS element to the incident signal, $\mathbf{H}_{\rm TI} \in \mathbb{C}^{L \times N_t}$ and $\mathbf{H}_{\rm RI}\in \mathbb{C}^{N_r \times L}$ denote channel matrix between the Tx and the RIS and between the RIS and Rx, respectively.
In mmWave channels, a geometrical channel model is widely applied.
Since the channel in mmWave systems has limited scattering, we can simplify the channel by assuming each cluster only contributes one path to the channel matrix \cite{ding2017ssm}.
Without loss of generality, we adopt the narrow band discrete physical channel model, which is an equivalent simplified version of the geometrical channel.
Here,  the $\mathbf{H}_{\rm TI}$ and $\mathbf{H}_{\rm RI}$ can be characterized as\footnote{If there exist the LoS path between the Tx-RIS or RIS-Rx channels, the spatial domain modulation gain will disappear, and the spectral efficiency will decrease.
In other words, if the transmitted beam and the reflected beam do not pass through the scatterers, the system degenerates into a traditional single-stream transmission communication system without SSM, and the spectral efficiency will be reduced.}
\begin{equation}\label{hh3}
\begin{aligned}
\mathbf{H}_{\rm TI} &= \sum_{n=1}^N g_n \boldsymbol{\alpha}_r(\vartheta^r_n,\eta^r_n)\boldsymbol{a}_t^H(\theta_n^t), \\ \mathbf{H}_{\rm RI} &= \sum_{m=1}^M h_m \boldsymbol{a}_r(\theta_m^r)\boldsymbol{\alpha}_t^H(\vartheta^t_m,\eta^t_m),
\end{aligned}
\end{equation}
where $N$ and $M$ are the numbers of candidate scatterers in the Tx-RIS and RIS-Rx channels, respectively; $g_n$ and $h_m$ stand for the complex gains of the $n$-th and $m$-th scatterers in the Tx-RIS and RIS-Rx channels, respectively; $g_n \sim \mathcal{CN}(0,1)$ and $h_m \sim \mathcal{CN}(0,1)$ represent the complex gains of the $n$-th and $m$-th scatterers from the Tx-RIS and RIS-Rx channels, respectively.
The steering vectors of the Rx and Tx with ULAs can be respectively calculated as
\begin{equation}
\begin{aligned}
&\boldsymbol{a}_r(\theta_m^r) = \frac{1}{\sqrt{N_r}}[1,e^{j\frac{2\pi d_r}{\lambda}\sin(\theta_m^r)},\cdots,
e^{j\frac{2\pi d_r}{\lambda}(N_r-1)\sin(\theta_m^r)}]^T, \\
&\boldsymbol{a}_t(\theta_n^t) = \frac{1}{\sqrt{N_t}}[1,e^{j\frac{2\pi d_t}{\lambda}\sin(\theta_n^t)},\cdots,
e^{j\frac{2\pi d_t}{\lambda}(N_t-1)\sin(\theta_n^t)}]^T,
\end{aligned}
\end{equation}
where $\lambda$ means the carrier wavelength. $\theta_m^r$ and $\theta_n^t$ stand for the angles of arrive and departure (AoA/AoD) for the $m$-th and $n$-th scatterers of channels $\mathbf{H}_{\rm RI}$ and $\mathbf{H}_{\rm TI}$, respectively.
Also, $d_r$ and $d_t$ denote the antenna spacings at the Rx and Tx sides, respectively.
In addition, the RIS is a uniform planar array (UPA) with $L=L_{\rm h}\times L_{\rm v}$ reconfigurable passive components,
where $L_{\rm h}$ and $L_{\rm v}$ represent the number of rows and columns of RIS, respectively.
Here, the receivie beam $\boldsymbol{\alpha}_r(\vartheta^r,\eta^r)$ and transmit beam $\boldsymbol{\alpha}_t(\vartheta^t,\eta^t)$ can be respectively characterized as
\begin{equation}
\begin{aligned}
\boldsymbol{\alpha}_r(\vartheta^r_n,\eta^r_n)&\!=\![1,
e^{j\frac{2\pi d}{\lambda}(\cos(\eta^r_n)\sin(\vartheta^r_n)+\sin(\eta^r_n))},\!\cdots, \\&
e^{j\frac{2\pi d}{\lambda}((L_{\rm h}-1)\cos(\eta^r_n)\sin(\vartheta^r_n)+(L_{\rm v}-1)\sin(\eta^r_n))}]^T,\\
\boldsymbol{\alpha}_t(\vartheta^t_m,\eta^t_m)&\!=\![1,
e^{j\frac{2\pi d}{\lambda}(\cos(\eta^t_m)\sin(\vartheta^t_m)+\sin(\eta^t_m))},\cdots,\\&
e^{j\frac{2\pi d}{\lambda}((L_{\rm h}-1)\cos(\eta^t_m)\sin(\vartheta^t_m)+(L_{\rm v}-1)\sin(\eta^t_m))}]^T,
\end{aligned}
\end{equation}
where $d$ denotes the interval of adjacent elements.
$\vartheta^r$ and $\eta^r$ are azimuth and elevation of AoA, respectively.
Meanwhile, $\vartheta^t$ and $\eta^t$ represent azimuth and elevation of AoD, respectively.

Combing (\ref{hh2}) and (\ref{hh3}), the channel in (\ref{hh1}) can be rewritten as
\begin{equation}\label{channel2}
\begin{aligned}
\mathbf{H} =& \sum_{m=1}^M \sum_{l=1}^L\sum_{n=1}^N h_m g_n\boldsymbol{a}_r(\theta_m^r)\boldsymbol{\alpha}_t^H(\vartheta^t_m,\eta^t_m) \\&\times e^{j\phi_l} \boldsymbol{\alpha}_r(\vartheta^r_n,\eta^r_n)\boldsymbol{a}_t^H(\theta_n^t).
\end{aligned}
\end{equation}
Let us define $\zeta^t_m = \arg\left[{\boldsymbol{\alpha}_t(\vartheta^t_m,\eta^t_m)}\right]$ and $\zeta^r_n = \arg\left[{\boldsymbol{\alpha}_r(\vartheta^r_m,\eta^r_m)}\right]$.
In this manner, (\ref{channel2}) can be recast as
\begin{equation}\label{hgsdf}
\mathbf{H} = \sum_{m=1}^M \sum_{l=1}^L\sum_{n=1}^N h_m g_n\boldsymbol{a}_r(\theta_m^r) e^{j\left(\phi_l+\zeta^t_m+\zeta^r_n\right)} \boldsymbol{a}_t^H(\theta_n^t).
\end{equation}
Note that we consider a large number of transmit and receive array antennas equipped at the Tx and Rx.
In this case, we use lemma 1 to give the corresponding characterization.
\begin{lemma}
With a large number of antenna elements at both the Tx and Rx sides, we have
\begin{equation}\label{orthog}
\begin{aligned}
&\boldsymbol{a}_t^H(\theta_{l_1})\boldsymbol{a}_t(\theta_{l_2})\approx \delta(l_1-l_2), \\
&\boldsymbol{a}_r^H(\theta_{l_1})\boldsymbol{a}_r(\theta_{l_2})\approx \delta(l_1-l_2).
\end{aligned}
\end{equation}
\end{lemma}
\emph{Proof:}
See Appendix A.

\begin{table}[t]
\small
\begin{tabular}{l}
\hline {\bf Algorithm 1} Suboptimal Detector of the RIS-DSSM Scheme \\
\hline
1: {\bf Input}: $\mathbf{H}$, $\boldsymbol{a}_r(\theta_m^r)$,
$\boldsymbol{a}_t^H(\theta_n^t), g_n, h_m, {P_s}, n_r, s_k, M, N, K, L$ \\
2: {\bf Output}: The detected $h_{\hat m}, g_{\hat n}, s_{\hat k}$ for bit demapping. \\
3:  {\bf for} ${\hat m}\leftarrow1: M$ {\bf do}\\
4:  $\quad ${\bf if} ${\hat m} == m$ {\bf then}\\
5:  $\quad \quad\mathbf{y}_{r}({\hat m})\leftarrow\boldsymbol{a}_r^H(\theta_m^r)\mathbf{H}\boldsymbol{a}_t(\theta_n^t)\sqrt{P_s}s_k+n_r$; \\
6:  $\quad ${\bf else} \\
7:  $\quad \quad\mathbf{y}_{r}({\hat m})\leftarrow n_r$; \\
8:  $\quad ${\bf end if} \\
9:   {\bf end for} \\
10: $\Delta_1\leftarrow$ $\infty$; \\
11: {\bf for} $m \leftarrow 1:M$ {\bf do}\\
12: $\quad $ {\bf for} $n \leftarrow 1:N$ {\bf do}\\
13: $\quad \quad$ {\bf for} $k \leftarrow 1:K$ {\bf do}\\
14:  $\quad \quad \quad r_1 \leftarrow$ $\boldsymbol{a}_r^H(\theta_m^r)\mathbf{H}\boldsymbol{a}_t(\theta_n^t)\sqrt{P_s}s_k$; \\
15: $\quad \quad \quad \quad$  $\Delta_2 \leftarrow$ $|\mathbf{y}_{r}({ m})-r_1 |$;\\
16:   $\quad \quad \quad$ {\bf if} $\Delta_2$ $<$ $\Delta_1$ {\bf then}\\
17:   $\quad \quad \quad \quad$ ${\hat m} \leftarrow m$;
   $\quad$ ${\hat n} \leftarrow n$;\\
18:   $\quad \quad \quad \quad  \quad$   ${\hat k} \leftarrow k$;
  $\quad $ $\Delta_1$ $\leftarrow$ $\Delta_2$;\\
19:   $\quad \quad \quad$ {\bf end if} \\
20: $\quad \quad$ {\bf end for}\\
21: $\quad $ {\bf end for} \\
22:  {\bf end for}\\
\hline
\end{tabular}
\end{table}
\subsection{RIS-DSSM Detection}
At the Rx, the received signal can be formulated as
\begin{equation}\label{eqt1}
\mathbf{y} = \sqrt{P_s}\mathbf{H}\boldsymbol{a}_t(\theta_n^t)s_k + \mathbf{n},
\end{equation}
Substituting  (\ref{hgsdf}) into (\ref{eqt1}), we have
\begin{equation}\label{eqt2}
\begin{aligned}
\mathbf{y} = &\sqrt{P_s}\sum_{m=1}^M \sum_{l=1}^L\sum_{n=1}^N h_m g_n\boldsymbol{a}_r(\theta_m^r) e^{j\left(\phi_l+\zeta^t_m+\zeta^r_n\right)} \\& \times\boldsymbol{a}_t^H(\theta_n^t)\boldsymbol{a}_t(\theta_n^t)s_k + \mathbf{n}.
\end{aligned}
\end{equation}
Resort to lemma 1, (\ref{eqt2}) can be updated as
\begin{equation}\label{eqt3}
\begin{aligned}
\mathbf{y} = &\sqrt{P_s}\sum_{m=1}^M \sum_{l=1}^L h_m g_n\boldsymbol{a}_r(\theta_m^r) e^{j\left(\phi_l+\zeta^t_m+\zeta^r_n\right)} s_k + \mathbf{n}.
\end{aligned}
\end{equation}
where $\mathbf{n}$ stands for the noise vector following $\mathcal{CN}(0,N_0\mathbf{I}_{N_r})$.
It is assumed that Rx knows perfect CSI through channel estimation,
each set of phase shifter networks connected to each RF chain is responsible for monitoring the beam from a specific scatterer.
In Fig. \ref{sys}, the received signal $\mathbf{y}$ is filtered by $M$ sets phase shifters.
If the number of received RF chains is greater than $M$, Rx has the ability to distinguish the beam from a specific scatterer to achieve the correct decoding of spatial domain information. On the contrary, the number of RF chains is insufficient to detect the beam directions of all candidate scatterers, which will lead to a deterioration in the detected performance.
Consequently, the weights of the minimum phase shifter network at Rx can be characterized as
\begin{equation}
\mathbf{R} = [\boldsymbol{a}_r(\theta_{1}^r),\ldots, \boldsymbol{a}_r(\theta_{m}^r),\ldots, \boldsymbol{a}_r(\theta_{M}^r)].
\end{equation}
After the RF chains, the received signal can be rewritten as
\begin{equation}\label{eqt4}
\begin{aligned}
\mathbf{y}_r = \mathbf{R}^H\mathbf{y}.
\end{aligned}
\end{equation}
Based on (\ref{eqt3}), the (\ref{eqt4}) can be further given as
\begin{equation}
\begin{aligned}
\mathbf{y}_r =& \sqrt{P_s}\sum_{m=1}^M \sum_{l=1}^L h_m g_n\boldsymbol{a}^H_r(\theta_{m}^r)\boldsymbol{a}_r(\theta_m^r) \\& \times e^{j\left(\phi_l+\zeta^t_m+\zeta^r_n\right)} s_k + \mathbf{R}^H\mathbf{n}.
\end{aligned}
\end{equation}
By utilizing lemma 1, we have
\begin{equation}
\begin{aligned}
\mathbf{y}_r =& \sqrt{P_s}\sum_{l=1}^L h_m g_n  e^{j\left(\phi_l+\zeta^t_m+\zeta^r_n\right)} s_k + \mathbf{R}^H\mathbf{n}.
\end{aligned}
\end{equation}
The goal of RIS is to offset the transmit and receive beam by adjusting the phase shift, which can be set as $\phi_l + \zeta^t_m+\zeta^r_n=0$. As such, we have
\begin{equation}
\begin{aligned}
\mathbf{y}_r  =\sqrt{P_s}L h_m g_n   s_k + \mathbf{R}^H\mathbf{n}.
\end{aligned}
\end{equation}
\subsubsection{Optimal Detector}
The {ML} detection algrithm of the proposed RIS-DSSM scheme can be formulated as
\begin{equation}\label{rese}
[\hat{n}, \hat{m}, \hat{k}]=\underset{\substack{
n \in\left\{1, \ldots, N\right\}
\\ m \in\left\{1, \ldots, M\right\} \\k \in\left\{1, \ldots, K\right\} \\}}{\arg \min }\left\|\mathbf{y}_r-\boldsymbol{a}_r^H(\theta_m^r)\mathbf{H}\boldsymbol{a}_t(\theta_n^t)\sqrt{P_s}s_k\right\|^2,
\end{equation}
where $\hat{n}$, $\hat{m}$, $\hat{k}$ stand for the indices of detected scatterers and symbol.
\subsubsection{Suboptimal Detector}
Since ML detection requires traversing all possibilities, this leads to the excessive complexity of decoding detection. To address this issue, we present a suboptimal detection algorithm.
To be specific, we perform energy-based detection of the receiving beam index as
\begin{equation}\label{subp1}
\hat{m}=\underset{m \in\left\{1,2, \ldots, M\right\}}{\arg \max }\left|\mathbf{y}_r(m)\right|^2,
\end{equation}
where the receiving beam from $\hat m$-th scatterer with the maximum power is selected.
Here, the signal at the $\hat m$-th RF chain can be characterized as
\begin{equation}\label{subp2}
\begin{aligned}
\mathbf{y}_r(\hat m)& = \boldsymbol{a}_r^H(\theta_{\hat m}^r)\mathbf{y}\\&=
\begin{cases}
\boldsymbol{a}_r^H(\theta_{\hat m}^r)\mathbf{n},  & \hat{m} \neq m,\\
\boldsymbol{a}_r^H(\theta_m^r)\mathbf{H}\boldsymbol{a}_t(\theta_n^t)\sqrt{P_s}s_k+\boldsymbol{a}_r^H(\theta_m^r)\mathbf{n}, & \hat{m} = m.
\end{cases}
\end{aligned}
\end{equation}
Assuming the scatterer decision in the RIS-Rx channel is correct, the signal from the $m$-th scatterer can be expressed as
\begin{equation}\label{subp3}
\mathbf{y}_r(m)=\boldsymbol{a}_r^H(\theta_m^r)\mathbf{H}\boldsymbol{a}_t(\theta_n^t)\sqrt{P_s}s_k + n_r,
\end{equation}
where noise term $n_r =\boldsymbol{a}_r^H(\theta_m^r)\mathbf{n}$ follows $\mathcal{CN}(0,N_0)$.
Accordingly, the transmit signals can be jointly detected as
\begin{equation}\label{subp4}
[\hat{n}, \hat{m}, \hat{k}]=\underset{\substack{
n \in\left\{1, \ldots, N\right\}
\\m \in\left\{1, \ldots, M\right\}
\\k \in\left\{1, \ldots, K\right\} \\}}{\arg \min }\left|\mathbf{y}_r(m)-\boldsymbol{a}_r^H(\theta_m^r)\mathbf{H}\boldsymbol{a}_t(\theta_n^t)\sqrt{P_s}s_k\right|^2.
\end{equation}

\section{Performance Analysis}

Based on suboptimal detector provided in (\ref{subp4}), the CPEP of the proposed RIS-DSSM can be calculated as
\begin{equation}\label{cpepcd0}
\begin{aligned}
{P}_b
=&\Pr\left(|\mathbf{y}_r(m)-\boldsymbol{a}_r^H(\theta_m^r)\mathbf{H}\boldsymbol{a}_t(\theta_n^t)\sqrt{P_s}s_k|^2
\right.\\&\left.>|\mathbf{y}_r(\hat m)-\boldsymbol{a}_r^H(\theta_{\hat m}^r)\mathbf{H}\boldsymbol{a}_t(\theta_{\hat{n}}^t)
\sqrt{P_s}s_{\hat k}|^2\right).
\end{aligned}
\end{equation}
For subsequent analysis, we treat both sides of the inequality sign in (\ref{cpepcd0}) according to (\ref{subp1})-(\ref{subp4}) to obtain
\begin{equation}\label{eqm1}
|\mathbf{y}_r(m)-\boldsymbol{a}_r^H(\theta_m^r)\mathbf{H}\boldsymbol{a}_t(\theta_n^t)\sqrt{P_s}s_k|^2
=|n_r|^2
\end{equation}
and
\begin{equation}\label{eqm2}
\begin{aligned}
&|\mathbf{y}_r(\hat m)-\boldsymbol{a}_r^H(\theta_{\hat m}^r)\mathbf{H}\boldsymbol{a}_t(\theta_{\hat{n}}^t)
\sqrt{P_s}s_{\hat k}|^2\\=&
\begin{cases}
|n_r-\sqrt{P_s}Lh_{\hat m}g_{\hat n}s_{\hat k}|^2, & \hat{m} \neq m \\
|\sqrt{P_s}Lh_{m}(g_ns_k-g_{\hat{n}}s_{\hat k})+n_r|^2, & \hat{m} = m.
\end{cases}
\end{aligned}
\end{equation}
After that, we will discuss the CPEP into two cases $\hat m =m$ and $\hat m \neq m$ separately.
\subsection{$\hat{m} = m$}
In this case, the beam decision at the RIS-Rx channel is correct, the CPEP  can be derived by substituing (\ref{eqm1}) and (\ref{eqm2}) into (\ref{cpepcd0}) as
\begin{equation}\label{cpepcd}
\begin{aligned}
{P}_b
=&\Pr\left(|n_r|^2>|\sqrt{P_s}Lh_{m}(g_ns_k-g_{\hat{n}}s_{\hat k})+n_r|^2\right)\\
=&\Pr\left(-2\Re\{N_0\sqrt{P_s}Lh_{m}(g_ns_k-g_{\hat{n}}s_{\hat k})\}\right.\\&\left.>|\sqrt{P_s}Lh_{m}(g_ns_k-g_{\hat{n}}s_{\hat k})|^2\right)\\
=&\Pr\left(-2\Re\{N_0\sqrt{P_s}Lh_{m}(g_ns_k-g_{\hat{n}}s_{\hat k})\}\right.\\&\left.-|\sqrt{P_s}Lh_{m}(g_ns_k-g_{\hat{n}}s_{\hat k})|^2>0\right)\\
=&\Pr\left(J>0\right),
\end{aligned}
\end{equation}
where $J$ follows $\mathcal{N}(\mu_J,\sigma_J^2)$, where the expectation and variance of $J$ are
$\mu_J = -|\sqrt{P_s}Lh_{m}(g_ns_k-g_{\hat{n}}s_{\hat k})|^2$ and $\sigma_J^2=2N_0\sqrt{P_s}Lh_{m}(g_ns_k-g_{\hat{n}}s_{\hat k})$, respectively.
As such, the (\ref{cpepcd}) can be further characterized as
\begin{equation}\label{pr0}
\begin{aligned}
&{P}_b
=Q\left(\sqrt{\frac{\mu_J^2}{\sigma_J^2}}\right)
=Q\left(\sqrt{\frac{P_s|Lh_{m}(g_ns_k-g_{\hat{n}}s_{\hat k})|^2}{2N_0}}\right).
\end{aligned}
\end{equation}
Since $g_n$ and $s_k$ are coupled with each other, it is difficult to solve them directly. Consequently, we address this term by using {\bf Lemma 2} as follows:
\begin{lemma}
To tackle this issue, let us define $\eta=|g_ns_k-g_{\hat n}s_{\hat k}|^2$, the average $\eta$ can be characterized as
\begin{equation}
\bar{\eta}
=
\begin{cases}
|s_{k}-{s}_{\hat{k}}|^2,& \text{$\hat{n}$ $=$ $n$}\\
|s_{k}|^2+|{s}_{\hat{k}}|^2,& \text{$\hat{n}$ $\neq$ $n$}.
\end{cases}
\end{equation}
\end{lemma}
\emph{Proof:} See in Appendix B.
$\hfill\blacksquare$

To obtain the closed-form expression of the CPEP, we need to perform an expectation operation on channel variables, which can be expressed as
\begin{equation}\label{pdfk001}
\begin{aligned}
&\bar{P}_b =\int_0^\infty Q\left(\sqrt{\frac{\rho L^2\bar\eta x }{2}}\right)f(x)dx,
\end{aligned}
\end{equation}
where $x=|g_nh_m|^2$, and $\rho={P_s}/{N_0}$ denotes the average SNR. Then, we rely on the following theorem to obtion the PDF of the variable $x$.
\begin{theorem}
The PDF of $x=|g_nh_m|^2$ can be evaluated as
\begin{equation}\label{pdfk0}
f(x)=2K_0(2\sqrt{x}).
\end{equation}
\end{theorem}
\emph{Proof:} See in Appendix C.
$\hfill\blacksquare$

On the basis of {\bf Theorem 1}, we subsitute this into the (\ref{pdfk001}). As such, the (\ref{pdfk001}) can be formulated as
\begin{equation}\label{pepee}
\begin{aligned}
&\bar{P}_b
=2\int_0^\infty Q\left(\sqrt{\frac{\rho L^2\bar\eta x }{2}}\right)K_0(2\sqrt{x})dx.
\end{aligned}
\end{equation}

In the following, we derive the PEP expression of the proposed RIS-DSSM scheme via the two methods.
\subsubsection{\bf Method 1}
In this case, we derive the exact analytical expression of PEP depending on the numerical integral approach.
To be specific, we substitute
$
Q(x) = \frac{1}{\pi}\int_0^\infty\exp\left(-\frac{x^2}{2\sin^2\theta}\right)d\theta
$
into (\ref{pepee}), the (\ref{pepee}) can be rewritten as
\begin{equation}\label{ddf0s}
\begin{aligned}
&\bar{P}_b
=\frac{2}{\pi}\int_0^\infty\int_0^{\frac{\pi}{2}}\exp\left(-\frac{{\rho L^2\bar{\eta} x}}{4\sin^2\theta}\right)K_0(2\sqrt{x})d\theta dx.
\end{aligned}
\end{equation}
By exchanging the order of integral variables $\theta$ and $x$, (\ref{ddf0s}) can be recast as
\begin{equation}\label{ddfs}
\begin{aligned}
&\bar{P}_b
=\frac{2}{\pi}\int_0^{\frac{\pi}{2}}{\int_0^\infty\exp\left(-\frac{{\rho L^2\bar{\eta} x}}{4\sin^2\theta}\right)K_0(2\sqrt{x})dx}d\theta.
\end{aligned}
\end{equation}
To address the inner integral centrally, we let
\begin{equation}\label{ddfs1}
\Lambda=\int_0^\infty\exp\left(-\frac{{\rho L^2\bar{\eta} x}}{4\sin^2\theta}\right)K_0(2\sqrt{x})dx.
\end{equation}
Since the procedure for solving for (\ref{ddfs1}) is tedious, we turn to \cite{jef2007book}
\begin{equation}\label{xxsd}
\begin{aligned}
&\int_0^\infty \exp(-\alpha x)K_{2v}(2\sqrt{\beta x})dx \\= & \frac{\exp\left(\frac{\beta}{2\alpha}\right)}{2\sqrt{\alpha\beta}}\Gamma(v+1)\Gamma(1-v)W_{-\frac{1}{2},v}\left(\frac{\beta}{\alpha}\right).
\end{aligned}
\end{equation}
Afterwards, we substitute (\ref{xxsd}) into (\ref{ddfs1}), the $\Lambda$ can be re-expressed as
\begin{equation}\label{lambb}
\begin{aligned}
\Lambda
&= \frac{\exp\left(\frac{2\sin^2\theta}{\rho L^2\bar{\eta}}\right)}{\sqrt{\frac{{\rho L^2\bar{\eta} }}{\sin^2\theta}}}\Gamma(1)\Gamma(1)W_{-\frac{1}{2},0}\left(\frac{4\sin^2\theta}{{\rho L^2\bar{\eta}}}\right)\\&=\frac{\exp\left(\frac{2\sin^2\theta}{\rho L^2\bar{\eta}}\right)\sin\theta}{L\sqrt{{{\rho \bar{\eta} }}}}W_{-\frac{1}{2},0}\left(\frac{4\sin^2\theta}{{\rho L^2\bar{\eta}}}\right).
\end{aligned}
\end{equation}
By combining (\ref{lambb}) and (\ref{ddfs}), the UPEP can be obtained as
\begin{equation}\label{eq31e}
\begin{aligned}
\bar{P}_b =&\frac{2}{\pi}\int_0^{\frac{\pi}{2}}\frac{\exp\left(\frac{2\sin^2\theta}{\rho L^2\bar{\eta}}\right)\sin\theta}{L\sqrt{{{\rho \bar{\eta} }}}}W_{-\frac{1}{2},0}\left(\frac{4\sin^2\theta}{{\rho L^2\bar{\eta}}}\right)d\theta \\=&\frac{2}{\pi{L\sqrt{{{\rho \bar{\eta} }}}}}\int_0^{\frac{\pi}{2}}{\exp\left(\frac{2\sin^2\theta}{\rho L^2\bar{\eta}}\right)}W_{-\frac{1}{2},0}\left(\frac{4\sin^2\theta}{{\rho L^2\bar{\eta}}}\right)\sin\theta d\theta.
\end{aligned}
\end{equation}
\begin{remark}
Since (\ref{eq31e}) contains the Whittaker function, it does not satisfy the form of the closed-form expression.
Recall  that (\ref{eq31e}) is the integral variable $\theta \in \left[0,\frac{\pi}{2}\right]$,  the upper bound in this case can be given as:
\begin{equation}\label{exact1}
\begin{aligned}
&\bar{P}_b
\leq \frac{2}{\pi{L\sqrt{{{\rho \bar{\eta} }}}}}{\exp\left(\frac{2}{\rho L^2\bar{\eta}}\right)}W_{-\frac{1}{2},0}\left(\frac{4}{{\rho L^2\bar{\eta}}}\right).
\end{aligned}
\end{equation}
Although (\ref{exact1}) satisfies the closure expression form, it is a loose upper bound of UPEP.
\end{remark}
To obtain a more accurate UPEP closed-form expression, we provide an alternative method 2 to deal with it.
\subsubsection{\bf Method 2}
In order to obtain the closed-form expression of (\ref{pepee}), it is necessary to find an equivalent form of $K_0(x)$ function to expand it. Fortunately, according to \cite{jef2007book}, we can obtain
\begin{equation}\label{K00}
K_0(x) = -\ln\left(\frac{x}{2}\right)I_0(x) + \sum_{v=0}^\infty\frac{x^{2v}}{2^{2v}(v!)^2}\psi(v+1).
\end{equation}
Further, the series expansion for the modified Bessel function $I_0(x)$ can be written as \cite{jef2007book}
\begin{equation}\label{K001}
I_0(x) = \sum_{v=0}^\infty\frac{(\frac{x}{2})^{2v}}{(v!)^2}.
\end{equation}
Substituting (\ref{K001}) into (\ref{K00}), we have
\begin{equation}\label{K100}
K_0(x) = \sum_{v=0}^\infty\frac{x^{2v}}{2^{2v}(v!)^2}\left[-\ln\left(\frac{x}{2}\right)+
\psi(v+1)\right].
\end{equation}
At this point, we substitute (\ref{K100}) into (\ref{pepee}), then the detailed derivation of $\bar P_b$ is given as follows:
\begin{equation}\label{pepe11}
\begin{aligned}
\bar{P}_b
= &2\int_0^\infty Q\left(\sqrt{\frac{\rho L^2\bar{\eta} x}{2}}\right) \\&\times \left(\sum_{v=0}^\infty\frac{x^{v}}{(v!)^2}\left[-\ln\left(\sqrt{x}\right)+
\psi(v+1)\right]\right)dx \\
=& 2\sum_{k=0}^\infty\frac{1}{(k!)^2}\left[\int_0^\infty Q\left(\sqrt{\frac{\rho L^2\bar{\eta} x}{2}}\right) \right.\\&\left.\times \left({x^{k}}\left[\ln\left(\sqrt{x}\right)+
\psi(k+1)\right]\right)dx \right]\\
=& -2\sum_{v=0}^\infty\frac{1}{(v!)^2}{\int_0^\infty Q\left(\sqrt{\frac{\rho L^2\bar{\eta} x}{2}}\right) {x^{v}}\ln\left(\sqrt{x}\right)dx} \\&+2\sum_{v=0}^\infty\frac{\psi(v+1)}{(v!)^2}\underbrace{\int_0^\infty Q\left(\sqrt{\frac{\rho L^2\bar{\eta} x}{2}}\right) {x^{v}}dx }_{A_2}\\
=& -\sum_{v=0}^\infty\frac{1}{(v!)^2}\underbrace{\int_0^\infty Q\left(\sqrt{\frac{\rho L^2\bar{\eta} x}{2}}\right) {x^{v}}\ln\left({x}\right)dx}_{A_1} \\&+2\sum_{v=0}^\infty\frac{\psi(v+1)}{(v!)^2}{A_2}.
\end{aligned}
\end{equation}
Due to the complexity of the (\ref{pepe11}), it is challenging to solve it directly.
In this case,
we subsequently handle $A_1$ and $A_2$ in terms of {\bf Theorem 2} and {\bf Theorem 3}, respectively.
\begin{theorem}
The closed-form expression of $A_1$ can be given by
\begin{equation}\label{theorm22}
\begin{aligned}
A_1 =& \frac{v!}{2}\left(\frac{4}{{\rho L^2 \bar{\eta}}}\right)^{v+1}\frac{ (2v+1)!!}{(2v+2)!!}\\& \times\left[
\ln\left(\frac{4}{\rho L^2 \bar{\eta}}\right) + \psi\left(v+\frac{3}{2}\right)-\frac{1}{v+1}
\right].
\end{aligned}
\end{equation}
\end{theorem}
\emph{Proof:} See in Appendix D.
$\hfill\blacksquare$

\begin{theorem}
The closed-form expression for $A_2$ can be characterized as
\begin{equation}\label{theorm23}
A_2 = \frac{v!}{2}\left(\frac{4}{\rho L^2 \bar{\eta}}\right)^{v+1}\frac{(2v+1)!!}{(2v+2)!!}.
\end{equation}
\end{theorem}
\emph{Proof:} See in Appendix E.
$\hfill\blacksquare$

By combing (\ref{theorm22}) and (\ref{theorm23}), the UPEP in (\ref{pepe11}) can be further calculated as
\begin{equation}\label{close1}
\begin{aligned}
\bar{P}_b
= &\sum_{v=0}^\infty\frac{2}{(v!)^2}\left[-\frac{1}{2}{A_1} +\psi(v+1){A_2}\right]\\
=& \sum_{v=0}^\infty\frac{A_2}{(v!)^2}
\left[-\ln\left(\frac{4}{\rho L^2 \bar{\eta}}\right) - \psi\left(v+\frac{3}{2}\right)\right.\\&\left.+\frac{1}{v+1}
 +2\psi(v+1)\right]\\
 =& \sum_{v=1}^\infty
\frac{1}{2(v-1)!}\left(\frac{4}{\rho L^2 \bar{\eta}}\right)^v\frac{(2v-1)!!}{(2v)!!}\\&\times
\left[-\ln\left(\frac{4}{\rho L^2 \bar{\eta}}\right) - \psi\left(v+\frac{1}{2}\right)+\frac{1}{v}
 +2\psi(v)\right].
\end{aligned}
\end{equation}

\subsection{$\hat{m} \neq m$}
This situation indicates that the beam direction from RIS to Rx is incorrectly decided.
Substituing (\ref{eqm1}) and (\ref{eqm2}) into (\ref{cpepcd0}), we have
\begin{equation}
\begin{aligned}
&{P}_b
={P}_r\left(|\boldsymbol{a}_r^H(\theta_m^r)\mathbf{n}|^2>|\boldsymbol{a}_r^H(\theta_{\hat m}^r)\mathbf{n}-\sqrt{P_s}Lh_{\hat m}g_{\hat n}s_{\hat k}|^2\right).
\end{aligned}
\end{equation}
Without loss of generality, let us define
\begin{equation}
w_1= |\boldsymbol{a}_r^H(\theta_m^r)\mathbf{n}|^2, \ \ w_2=|\boldsymbol{a}_r^H(\theta_{\hat m}^r)\mathbf{n}-\sqrt{P_s}Lh_{\hat m}g_{\hat n}s_{\hat k}|^2,
\end{equation}
where
$
\boldsymbol{a}_r^H(\theta_l^r)\mathbf{n} \sim \mathcal{CN}(0,N_0)$ and
$\left(\boldsymbol{a}_r^H(\theta_{\hat m}^r)\mathbf{n}-\sqrt{P_s}Lh_{\hat m}g_{\hat n}s_{\hat k}\right) \sim \mathcal{CN}(-\sqrt{P_s}Lh_{\hat m}g_{\hat n}s_{\hat{k}},N_0).
$
It can be easily found that ${w_1}$ is the central Chi-squared random variable with two degrees of freedom (DoFs),
while ${w_2}$ means the non-central Chi-squared random variable with two DoF.
Furthermore, let us make ${x_1 = \frac{w_1}{N_0/2}}$ and
$x_2 = \frac{w_2}{N_0/2}$.
At this moment, the central Chi-square distribution $x_1$ can be described as \cite{kopka}
\begin{equation}\label{np2}
f_1(x_1)=\frac{1}{2}\exp\left(-\frac{x_1}{2}\right).
\end{equation}
As a result, $\bar P_b$ can be calculated as
\begin{equation}\label{np3}
\begin{aligned}
 {P}_b
&=\frac{1}{2}\int_0^\infty f_2(x_2)\left(\int_{x_2}^\infty \exp\left(-\frac{x_1}{2}\right) dx_1\right) d x_2\\&
=\int_0^\infty f_2(x_2) \exp\left(-\frac{x_2}{2}\right) d x_2.
\end{aligned}
\end{equation}
It is observed that (\ref{np3}) satisfies the form of {moment generating function (MGF)}.
To address this issue, we resort to the solution provided by \cite{kopka}. Accordingly, the
MGF of noncentral Chi-square distribution with two DoFs can be evaluated as
\begin{equation}\label{mgf}
M_s(x_2) = \left(\frac{1}{1-2s\sigma^2}\right)\exp\left(\frac{s\mu^2}{1-2s\sigma^2}\right).
\end{equation}
Substituting (\ref{mgf}) into (\ref{np3}), we get
\begin{equation}
\begin{aligned}
{P}_b
=\left(\frac{1}{1+\sigma^2}\right)\exp\left(-\frac{\mu^2}{2(1+\sigma^2)}\right).
\end{aligned}
\end{equation}
For $x_2$, we have $\sigma^2=N_0$ and $\mu = -\sqrt{P_s}Lh_{\hat m}g_{\hat n}s_{\hat k}$. Consequently, the
CPEP can be calculated as
\begin{equation}\label{npep6}
\begin{aligned}
{P}_b
=&\frac{1}{2}{\exp\left(-\frac{\rho L^2|h_{\hat{m}}g_{\hat n}|^2|s_{\hat{k}}|^2}{2}\right)}
=\frac{1}{2}{\exp\left(-\frac{\rho L^2|s_{\hat{k}}|^2}{2}x\right)}.
\end{aligned}
\end{equation}
To eliminate the variables in (\ref{npep6}) with respect to $h_{\hat m}g_{\hat n}$, we require an expectation operation into the CPEP-based derivation of the UPEP expression.
In the following, the UPEP expressions in this context are derived by two methods.
\subsubsection{\bf Method 1}
In this case, we derive the exact analytical expression of UPEP via numerical integration method.
\begin{theorem}
In this $\hat m \neq m$ case, the UPEP of RIS-DSSM scheme can be given by
\begin{equation}\label{eq39}
\begin{aligned}
\bar{P}_b
&=\frac{\exp\left(\frac{1}{2\alpha}\right)}{2\sqrt{\alpha}} W_{-\frac{1}{2},0}\left(\frac{1}{\alpha}\right),
\end{aligned}
\end{equation}
where $\alpha = \frac{\rho L^2 |s_{\hat k}|^2}{2}$.

{Proof:}
By combing (\ref{pdfk0}) and (\ref{npep6}), the UPEP can be calculated as
\begin{equation}\label{eq40}
\begin{aligned}
\bar{P}_b&=\int_{0}^\infty K_0(2\sqrt{x})\exp\left(-\frac{\rho L^2 |s_{\hat{k}}|^2}{2}x\right)dx.
\end{aligned}
\end{equation}
Resort to (\ref{xxsd}), we can rewrite (\ref{eq40}) as
\begin{equation}\label{eq3444}
\begin{aligned}
\bar{P}_b
&=\frac{\exp\left(\frac{1}{2\alpha}\right)}{2\sqrt{\alpha}} \Gamma(v+1)\Gamma(1-v)W_{-\frac{1}{2},v}\left(\frac{1}{\alpha}\right).
\end{aligned}
\end{equation}
After some simplifications, we complete the proof.
\end{theorem}
Note that (\ref{eq39}) contains  Whittaker function, it is not closed-form expression, which limits the contribution of this work.
To tackle this issue, we resort to the following method.
\subsubsection{\bf Method 2}
To obtain the closed-form expression of UPEP, we can derive the UPEP under $\hat m \neq m$ scenario as
\begin{equation}\label{eq3569}
\begin{aligned}
\bar{P}_b
= &\frac{1}{2}\int_0^\infty \exp\left(-\frac{\rho L^2|s_{\hat k}|^2}{2}x\right)f(x)dx \\
= &\sum_{v=0}^\infty\frac{x^{v}}{(v!)^2}\int_0^\infty \exp\left(-\frac{\rho L^2|s_{\hat k}|^2}{2}x\right)\\&\times\left[-\frac{1}{2}\ln\left({x}\right)+
\psi(v+1)\right]dx \\
= &-\frac{1}{2}\sum_{v=0}^\infty\frac{1}{(v!)^2}\underbrace{\int_0^\infty x^{v}\ln\left({x}\right)\exp\left(-\frac{\rho L^2|s_{\hat k}|^2}{2}x\right)dx}_{C_1} \\&+ \sum_{v=0}^\infty\frac{\psi(v+1)}{(v!)^2}\underbrace{\int_0^\infty x^{v} \exp\left(-\frac{\rho L^2|s_{\hat k}|^2}{2}x\right)
dx}_{C_2},
\end{aligned}
\end{equation}
where $f(x)=2\sum_{v=0}^\infty\frac{x^{v}}{(v!)^2}\left[-\frac{1}{2}\ln\left({x}\right)+
\psi(v+1)\right]$ can be obtained via the (\ref{pdfk0}) and (\ref{K100}).

By applying the (\ref{eq51}), the $C_1$ in (\ref{eq3569}) can be obtained as
\begin{equation}\label{cc11}
C_1 = v!\left(\frac{2}{\rho L^2|s_{\hat k}|^2}\right)^{v+1}
\left[\psi(v+1)-\ln\left(\frac{\rho L^2|s_{\hat k}|^2}{2}\right)\right].
\end{equation}
According to \cite{jef2007book}, we get
\begin{equation}\label{eq35691}
\int_0^\infty x^p \exp(-\beta x^q) dx = \frac{\Gamma(\gamma)}{q\beta^\gamma}, \ \ \
\end{equation}
where $\gamma = \frac{p+1}{q}$.
By combining (\ref{eq35691}) and (\ref{eq3569}), then (\ref{eq3569}) can be re-represented as
\begin{equation}\label{cc12}
C_2 = v!\left(\frac{2}{\rho L^2|s_{\hat k}|^2}\right)^{v+1}.
\end{equation}
Replacing (\ref{cc11}) and (\ref{cc12}) into (\ref{eq3569}), we can obtain the UPEP as
\begin{equation}\label{close2}
\begin{aligned}
\bar{P}_b
= &-\frac{1}{2}\sum_{v=0}^\infty\frac{1}{v!}\left(\frac{2}{\rho L^2|s_{\hat k}|^2}\right)^{v+1}\left[\psi(v+1)-\ln\left(\frac{\rho L^2|s_{\hat k}|^2}{2}\right)\right]\\&+ \sum_{v=0}^\infty\frac{\psi(v+1)}{v!}\left(\frac{2}{\rho L^2|s_{\hat k}|^2}\right)^{v+1} \\
=& \frac{1}{2}\sum_{v=0}^\infty\frac{1}{v!}\left(\frac{2}{\rho L^2|s_{\hat k}|^2}\right)^{v+1}\left[
\ln\left(\frac{\rho L^2|s_{\hat k}|^2}{2}\right)+{\psi(v+1)} \right].
\end{aligned}
\end{equation}

\begin{figure}[t]
\centering
\includegraphics[width=7cm]{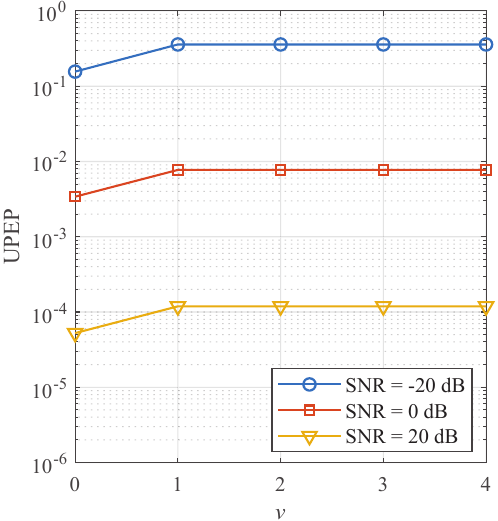}
\caption{\small{Converage analysis. }}
\vspace{-10pt}
\label{figcoverag}
\end{figure}
\subsection{Asymptotic UPEP}

In this subsection, we investigate the asymptotic expression of UPEP from (\ref{close1}) and (\ref{close2}).
It can be observed that the expressions (\ref{close1}) and (\ref{close2}) are summed by an infinite number of terms, which makes our computation very difficult.
To cope with this challenge, we select the term that plays a dominant role in UPEP as an asymptotic expression to characterize UPEP.
As shown in Fig. \ref{figcoverag}, we describe the relationship between the number of summation terms of (\ref{close1}) and (\ref{close2}) and the value of {UPEP}.
It can be observed that as $v$ increases, UPEP achieves convergence at $v = 1$.
In particular,  Fig. \ref{figcoverag} shows that there is a slight gap between the {UPEP} obtained when $v = 0$ and $v = 1$ due to the fact that in (\ref{close1}) the first term is obtained at $v=1$, while in (\ref{close2}) the first term is obtained from $v=0$.
Besides, Fig. \ref{figcoverag} shows that increasing the SNR leads to better UPEP performance.
As such, the first term of (\ref{close1}) and (\ref{close2}) can guarantee to represent the whole values of the (\ref{close1}) and (\ref{close2}), which validates that the correctness of the operation in (\ref{asy1}) and (\ref{asy2}), respectively.
As a result, we can obtain the approximating UPEP as follows:
\subsubsection{$\hat{m} = m$}
Recalling Fig. \ref{figcoverage}, we observe that most of the information is concentrated in the first item. Moreover, we aim to obtain the trend of ABEP changing with SNR. As such, the main term of the (\ref{close1}) can be obtained as
\begin{equation}\label{asy1}
\begin{aligned}
&\bar{P}_{\rm a}
 = \frac{1}{\rho L^2 \bar{\eta}}
\left[\ln\left(\frac{\rho L^2 \bar{\eta}}{4}\right) - \psi\left(\frac{3}{2}\right)+1
 +2\psi(1)\right].
\end{aligned}
\end{equation}

\subsubsection{$\hat{m} \neq m$}
Combing (\ref{close2}), the asymptotic PEP in this case can be evaluated as
\begin{equation}\label{asy2}
\begin{aligned}
&\bar{P}_{\rm a}
= \frac{1}{\rho L^2|s_{\hat k}|^2}\left[
\ln\left(\frac{\rho L^2|s_{\hat k}|^2}{2}\right)+{\psi(1)} \right].
\end{aligned}
\end{equation}

\subsection{Diversity Gain}
In this subsection, we derive the diversity order $\mathcal{D}$ of the proposed RIS-DSSM scheme based on the asymptotic UPEP expression.
It is worth noting that $\mathcal{D}$  is the high-SNR slope of the PEP determined from a figure plotted in a log-log scale.
In the following, the $\hat m = m$ and $\hat m \neq m$ cases are respectively derived as follows:
\subsubsection{$\hat{m} = m$}
In this case, the $\mathcal{D}$ can be evaluated as
\begin{equation}
\begin{aligned}
\mathcal{D}&=\lim\limits_{\rho \to \infty}-\frac{\log_2(\bar{P}_{\rm a} )}{\log_2\rho}=\lim\limits_{\rho \to \infty}\frac{\log_2\left(\frac{{\rho L^2 \bar{\eta}}}{\ln\left(\frac{\rho L^2\bar{\eta}}{4}\right)}\right)}{\log_2\rho}\\&=\lim\limits_{\rho \to \infty}\frac{\log_2\left({{\rho L^2 \bar{\eta}}}\right)}{\log_2\rho}=1.
\end{aligned}
\end{equation}
\subsubsection{$\hat{m} \neq m$}
In this case, the $\mathcal{D}$ can be calculated as
\begin{equation}
\begin{aligned}
\mathcal{D}&=\lim\limits_{\rho \to \infty}-\frac{\log_2(\bar{P}_{\rm a} )}{\log_2\rho}=\lim\limits_{\rho \to \infty}\frac{\log_2\left(\frac{\rho L^2|s_{\hat k}|^2}{\ln\left(\frac{\rho L^2|s_{\hat k}|^2}{2}\right)}\right)}{\log_2\rho}\\&=\lim\limits_{\rho \to \infty}\frac{\log_2\left(\rho L^2|s_{\hat k}|^2\right)}{\log_2\rho}=1.
\end{aligned}
\end{equation}

\subsection{ABEP}
In this subsection, we evaluate the error performance of the RIS-DSSM system, where $\log_2(NMK)$ bits are utilized for transmission. Note that the union upper bound of ABEP is given by
\begin{equation}\label{abep}
\begin{aligned}
{\rm ABEP} \leq & \sum_{n = 1}^N \sum_{{m}=1}^M \sum_{k=1}^{K}  \sum_{\hat{n} = 1}^N \sum_{{\hat{m}}=1}^M \sum_{\hat{k}=1}^{K}\frac{N([n,m,k] \to [\hat{n},\hat{m},\hat{k}])\bar{P}_b
}{{{NMK  \log _{2}(NMK)}}},
\end{aligned}
\end{equation}
where $N([n,m,k] \to [\hat{n},\hat{m},\hat{k}])$ represents the number of error bits between the original bits $[n,m,k]$ and the decision bits $[\hat{n},\hat{m},\hat{k}]$.

\section{Simulation and Analytical Results}
In this section, simulation and analytical results are presented to evaluate the RIS-DSSM performance for mmWave trnasmissions.
At first, the detection overhead and ABEP of the optimal and suboptimal detectors are evaluated, where the $N_t$ and $N_r$ are set as 32 for the simulations \cite{ding2017ssm}. Then, link-level simulations verify the correctness of the analytical derivations of the proposed RIS-DSSM scheme. Afterward, we evaluate the ABEP performance versus different parameters.
Finally, the proposed RIS-DSSM scheme is compared with the conventional SSM in terms of reliability.
All simulation results are randomly generated $1 \times 10^7$ channel realizations and then averaged.
The central frequency of the considered carrier in this paper is 28 GHz \cite{liu2021casc}.
\subsection{Comparing the Complexity of Suboptimal and Optimal Detectors}
In this subsection, we present the computational complexity of the detectors at the Rx for the RIS-DSSM schemes, which can be obtained by finding the number of real additions and real multiplications.
Noting that each complex multiplication requires 4 real multiplications and 2 real summations, while the square of the absolute value of the complex number requires 2 real multiplications and 1 real addition.
In the subsequent analysis, we compare the complexity of the suboptimal and optimal detectors for the RIS-DSSM scheme.

\begin{figure}
\centering
\subfigure[$N=2$]{
\begin{minipage}[b]{0.47\textwidth}
\includegraphics[width=7cm]{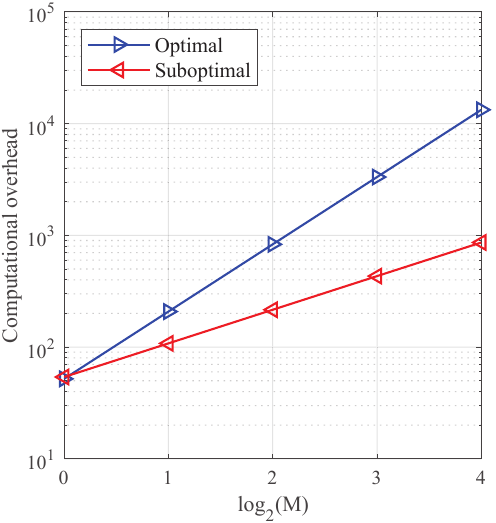}
\end{minipage}
}
\subfigure[$M=2$]{
\begin{minipage}[b]{0.47\textwidth}
\includegraphics[width=7cm]{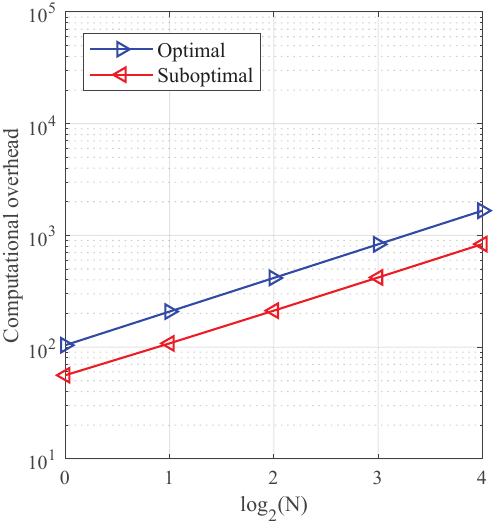}
\end{minipage}
}
\caption{\small{Comparison analysis of computational overhead.}}\label{complexity}
\end{figure}
To simplify the analysis, we apply (\ref{orthog}) to optimal and suboptimal detectors.
Recalling that suboptimal detection is composed of (\ref{subp1}) and (\ref{subp4}).
In terms of (\ref{subp1}), the real complexity of the required multiplications and summations are $2M$ and $M$, respectively.
For the suboptimal detector, the (\ref{subp4}) can be described as
\begin{equation}\label{eqq57}
[\hat{n}, \hat{m}, \hat{k}]=\underset{\substack{
n \in\left\{1, \ldots, N\right\}
\\ m \in\left\{1, \ldots, M\right\} \\k \in\left\{1, \ldots, K\right\} \\}}{\arg \min }\left|\mathbf{y}_r(m)-\sqrt{P_s}Lh_mg_ns_k\right|^2,
\end{equation}
where $h_{ m}g_{ n}s_{ k}$ can be calculated by the multiplication of three complex numbers with 8 real multiplications and 4 real summations.
Multiplication of $\sqrt{P_s}L$ accompanies 3 real multiplications, while substracting $\sqrt{P_s}Lh_{ m}g_{n}s_{k}$ from $\mathbf{y}_{r}({ m})$ requires 2 real summations.
The square of the absolute value of the calculated complex value is also taken into account, as well as the repetition of this process for each symbol and scatterer in the Tx-RIS and RIS-Rx channels, respectively
As such, the calculations of (\ref{subp1}) and (\ref{eqq57}) require $13MNK+2M$ multiplications and $7MNK+M$ summations.
On the other hand, for the optimal detector, the (\ref{rese}) based on (\ref{orthog}) can be obtained as
\begin{equation}\label{eqq58}
[\hat{n}, \hat{m}, \hat{k}]=\underset{\substack{
n \in\left\{1, \ldots, N\right\}
\\ m \in\left\{1, \ldots, M\right\} \\k \in\left\{1, \ldots, K\right\} \\}}{\arg \min }\sum_{m=1}^M\left|\mathbf{y}_r(m)-\sqrt{P_s}Lh_mg_ns_k\right|^2.
\end{equation}
It is worth noting that the complexity of the optimal detection algorithm is about $M$ times that of the suboptimal algorithm. The complexity of the optimal detection can be expressed as $13M^2NK$ multiplications and $7M^2NK$ summations.

The summation and multiplication operations have similar forms, without loss of generality, we only plot the complexity overhead of the multiplication operation in Fig. \ref{complexity}.
It is worth mentioning that unless otherwise stated, the parameters are set to $M=2, N=2, K=2$, and $L=100$.
As $M$ increases, the gap between optimal detection and suboptimal detection gradually widens shown in Fig. \ref{complexity}(a), while in Fig. \ref{complexity}(b), as $N$ increases, the gap between optimal detection and suboptimal detection remains stable.
This phenomenon can be explained by (\ref{eqq57}) and (\ref{eqq58}), respectively.
\begin{figure}
\centering
\subfigure[$N=2$]{
\begin{minipage}[b]{0.47\textwidth}
\includegraphics[width=7cm]{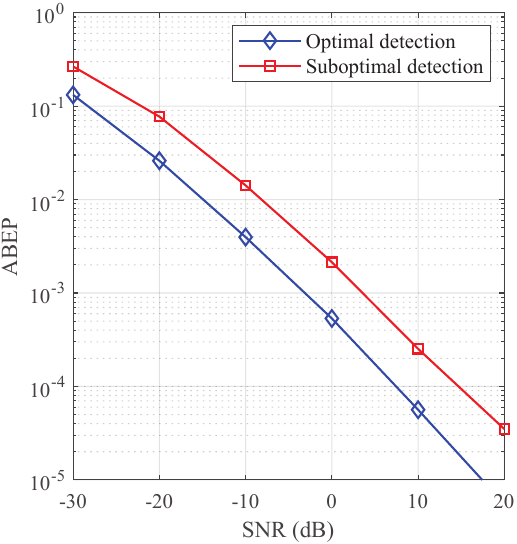}
\end{minipage}
}
\subfigure[$M=2$]{
\begin{minipage}[b]{0.47\textwidth}
\includegraphics[width=7cm]{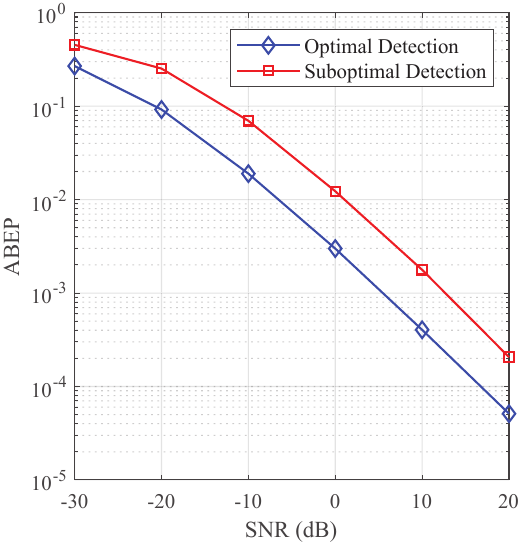}
\end{minipage}
}
\caption{\small{Comparison analysis of ABEP.}}\label{abepfi}
\end{figure}

\subsection{ABEP Comparison of Suboptimal and Optimal Detectors}
In {Fig. \ref{abepfi}}, we perform a comparative analysis of the ABEP performance of the optimal and suboptimal detection algorithms.
As expected, optimal detection exhibits better performance than suboptimal detection.
Unless otherwise stated, parameters are set to $M=2, N=2$, $K=2$, and $L=100$.
To be specific, when ABEP = $10^{-3}$,
the SNR required for the suboptimal algorithm in Fig. \ref{abepfi}(a) and Fig. \ref{abepfi}(b) is about 5 dB and 7 dB more than that of the optimal algorithm, respectively.
It is worth mentioning that the inferior performance of the suboptimal algorithm compared to the optimal algorithm is the cost of the detection complexity.

\begin{figure}
\centering
\subfigure[$L=100$]{
\begin{minipage}[b]{0.47\textwidth}
\includegraphics[width=7cm]{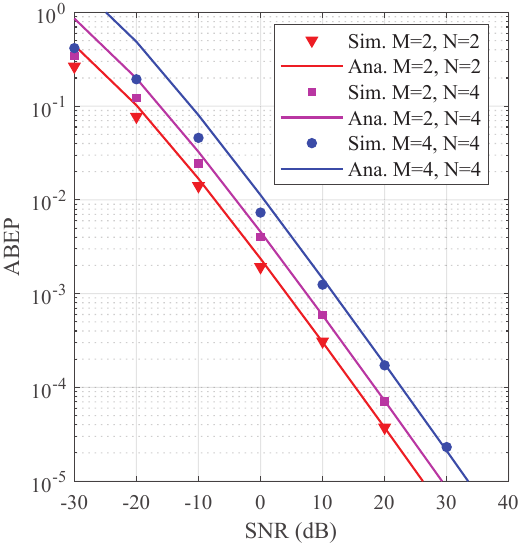}
\end{minipage}
}
\subfigure[{$L=64$}]{
\begin{minipage}[b]{0.47\textwidth}
\includegraphics[width=7cm]{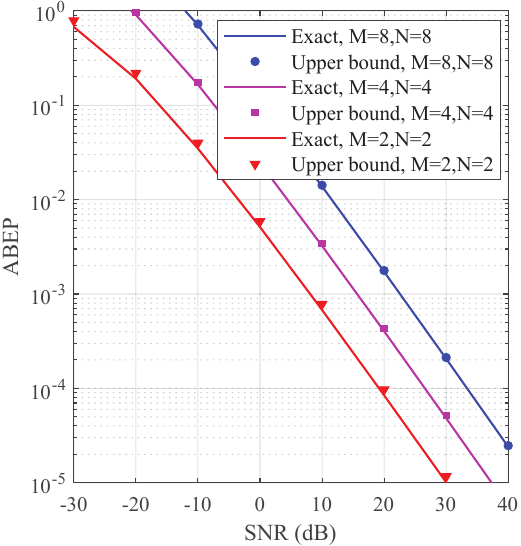}
\end{minipage}
}
\caption{\small{Verification of analytical derivation results. }}\label{figsim1}
\end{figure}
\subsection{Validation of Analytical Derivations}
In Fig. \ref{figsim1}(a), we illustrate the exactness of the union upper bound of ABEP for the proposed RIS-DSSM scheme, where the parameters are set as
with $M=2,N=2$, $M=2,N=4$, and $M=4,N=4$.
Without loss of generality, the BPSK signal for the symbol domain is employed.
From Fig. \ref{figsim1}(a), we see that the analytical and simulation results match closely at moderate to high SNRs, which illustrates the exactness of the analytical ABEP expressions of (\ref{eq31e}) and (\ref{eq39}).
Moreover, the ABEP performance of RIS-DSSM deteriorates as the number of modulated scatterers increases.
This is because involving more scatterers in modulation increases the number of constellation points available in the spatial domain, which reduces decoding success probability.
In an environment with abundant scatterers, we can select scatterers with better channel gain to participate in the modulation, while ignoring scatterers with poorer gain.
In this manner,
when more scatterers are participating in the modulation, the spectral efficiency increases.

In Fig. \ref{figsim1}(b), we validate the upper bound expressions (\ref{exact1}) of UPEP the versus the exact integral expression (\ref{eq31e}) of UPEP under different modulation orders with respect with the spatial domain.
It can be seen from Fig. \ref{figsim1}(b) that the ABEP based on upper bound expressions (\ref{exact1}) are closely matched by the ABEP based on exact UPEP integral expression.
This is because we divide the spatial domain into two parts, beam correct decoding and incorrect decoding, and the derived upper bound is in beam correct demodulation. The probability of beam incorrect decoding becomes larger as the modulation order in the spatial domain rises, and the percentage of this part of beam correct demodulation decreases. Hence, as the spatial domain modulation order rises, the difference between the obtained upper bound and the exact integral expression of ABEP is relatively small.

\begin{figure}[t]
\centering
\includegraphics[width=7cm]{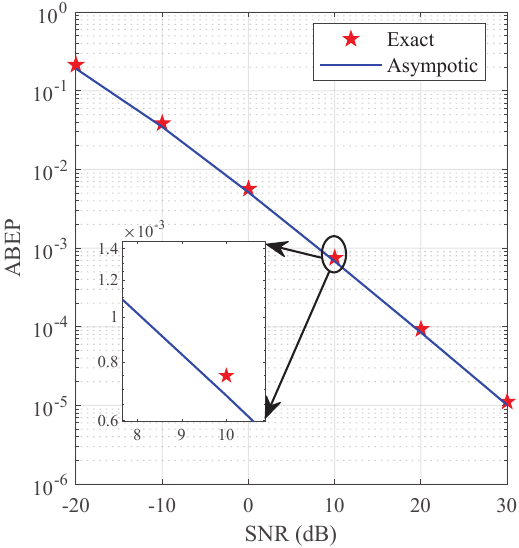}
\caption{\small{Verification of analytical derivation results. }}
\vspace{-10pt}
\label{figcoverage}
\end{figure}

In Fig. \ref{figcoverage}, we validate the accuracy of the derived closed-form asympotic expression with the exacat integral expression of ABEP, where closed-form expression is consist of
(\ref{asy1}) and (\ref{asy2}) and integral ABEP is consist of (\ref{eq31e}) and (\ref{eq39}).
The parameters are set as $M=2,N=2,K=2$, and $L=64$.
It can be seen from Fig. \ref{figcoverage} that, in the region of -20 dB to 30 dB, the closed-form asympotic expression and the integral expression are in good agreement. However, in the enlarged subfigure, we can find slightly lower values for the closed-form asympotic expression than the exact integral expression.
This is because we take the first term of (\ref{asy1}) and (\ref{asy2}) to approximate the whole expression, which is highlighted in Fig. \ref{figcoverag}.
To improve the accuracy, we can increase the number of selected terms of the closed-form expression.

\subsection{Performance Analysis of ABEP with Different Parameters}
\begin{figure}[t]
\centering
\subfigure[$M=2,N=2,L=100$]{
\begin{minipage}[b]{0.47\textwidth}
\includegraphics[width=7cm]{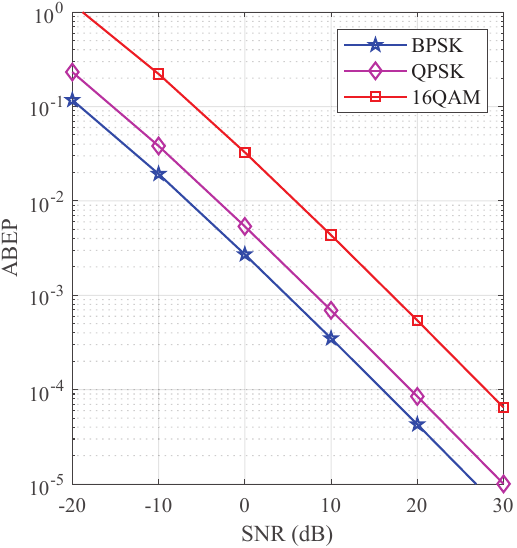}
\end{minipage}
}
\subfigure[$M=2,N=2,L=100$]{
\begin{minipage}[b]{0.47\textwidth}
\includegraphics[width=7cm]{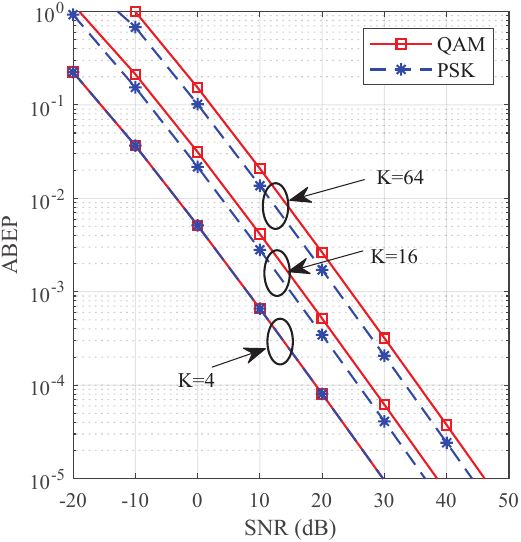}
\end{minipage}
}
\caption{\small{ABEP under different parameters. }}\label{figper}
\end{figure}

Fig. \ref{figper}(a) shows a comparison of the ABEPs for the RIS-DSSM system using BPSK, QPSK, and 16QAM signal constellations at data rates of 3 bit per channel use (bpcu), 4 bpcu, and 6 bpcu, respectively, where gray mapping is applied to all symbol modulations.
It can be observed from Fig. \ref{figper}(a) that these systems vary in performance.
To be specific, when ABEP is $10^{-4}$, systems with QPSK and 16QAM require approximately 3 dB and 9 dB higher SNRs than the BPSK system.
This is because, in the normalized symbol constellation diagram, the increasing modulation order leads to a denser distribution of constellation points, resulting in a smaller Euclidean distance between adjacent constellation points.

Fig. \ref{figper}(b) depicts the plot of the ABEP of the RIS-DSSM system versus the SNRs with $K$-ary PSK and QAM applied in the symbol domain.
It can be observed from Fig. \ref{figper}(b) that for the considered parameters with 4QAM and QPSK can obtain the same performance since the constellation diagrams of 4QAM and QPSK are consistent.
Further, it can also observed that at the same error rate, the higher the modulation order, the greater the required transmit SNR.
From this observation, we find that the system with 16PSK has a better performance than that of 16QAM. Meanwhile, the system with 64PSK performs better than that of 64QAM.
This is because (\ref{npep6})-based ABEP is only related to the amplitude of $ s_{\hat k}$, whereas in 16-order and 64-order modulation systems, the average amplitude of the PSK signal is large than that of the QAM signal.

\begin{figure}[t]
\centering
\includegraphics[width=7cm]{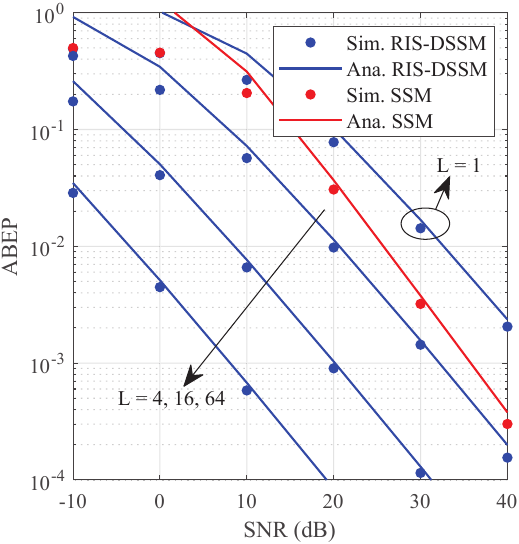}
\caption{\small{Comparison with traditional SSM scheme. }}
\vspace{-10pt}
\label{comSSM1}
\end{figure}
\subsection{Comparison to the Conventional SSM}
Fig. \ref{comSSM1} depicts the ABEP values of the SSM and RIS-DSSM schemes under different SNRs, where the number of RIS elements in the RIS-DSSM scheme is respectively taken as 1, 4, 16, and 64.
The simulation parameters of the RIS-DSSM scheme are configured as $M=2, N=2$, and $K=2$.
In contrast, the number of scatterers and the symbol field in the SSM scheme are set to 4 and BPSK, respectively.
As expected, the Monte Carlo simulation results are similar to the ABEP obtained via the closed-form expressions in (\ref{close1}) and (\ref{close2}). It can be seen that the performance of SSM dominates at $L=1$, while RIS-DSSM outperforms SSM at higher values of $L$.
The reason for this phenomenon is that when $L=1$, RIS cannot form a specific beam directed at the scatterer, which leads to a larger value of ABEP.
In contrast, when $L$ is larger, the beam directed to the scatterer is more refined and stronger in energy, thus obtaining a better system performance.
It also interesting to analyze the minimum $L$ for which the RIS-DSSM scheme outperforms the SSM scheme.
In this regard, when the receiving beam is demodulated correctly or not, we obtain the following two equations to determine the minimum $L$.
\begin{equation}\label{eq64}
\begin{aligned}
L^2- \frac{13}{ 24 }
\ln\left(\frac{\rho L^2 \bar{\eta}}{4}\right) +\frac{13}{ 24 }\left[\psi\left(\frac{3}{2}\right)-1
 -2\psi(1)\right]\leq0,
\end{aligned}
\end{equation}
\begin{equation}\label{eq65}
L^2-
\ln\left(\frac{\rho L^2|s_{\hat k}|^2}{2}\right)-{\psi(1)} \leq 0.
\end{equation}
The detailed derivation is presented in Appendix F.
However, due to the complexity of the formulas, it is not possible to obtain the closed-form solution directly. Nevertheless, we observe that the parameter affecting the performance of the two systems is not only related to the number of reflecting elements $L$, but also to the SNR $\rho$.

\section{Conclusions and Future Work}
In this paper, we conducted a study on the ABEP of a RIS-assisted DSSM system in mmWave MIMO systems. To decrease detection complexity, a suboptimal detection algorithm was presented and compared with the ML-based optimal detection algorithm.
Using the suboptimal detector, the exact integral ABEP and closed-form ABEP expressions were respectively obtained via the two different methods. Moreover, the upper bound and asymptotic expressions of ABEP on the RIS-DSSM scheme are provided.
Monte Carlo simulations were used to validate the correctness of the analytical derivations. Our results showed that the analytical upper bound and asymptotic ABEP keep close agreement with the exact integral results.
Furthermore, as the number of RIS elements increases, the RIS-DSSM scheme outperforms conventional SSM scheme in terms of reliability.
For future work, the impact of channel estimation errors on the performance of RIS-DSSM systems can be explored. In addition, evaluating the system performance from other metrics, such as ergodic capacity, and system throughput, is still an interesting and open research topic.

\begin{appendices}
\section{Proof of Lemma 1}
For the large number of the transmit antennas equipped at the Tx, the inner product of the two steering vectors can be written as
\begin{equation}\label{prlem1}
\begin{aligned}
&\boldsymbol{a}^H_t(\theta_n^t)\boldsymbol{a}_t(\theta_{n'}^t)
=\frac{1-\exp\left(j2\pi(\theta_{n'}^t-\theta_n^t)N_t\right)}{N_t-N_t\exp\left(j2\pi(\theta_{n'}^t-\theta_n^t)\right)}\\&
=\frac{\sin\left(\pi(\theta_{n'}^t-\theta_n^t)N_t\right)}{N_t\sin\left(\pi(\theta_{n'}^t-\theta_n^t)\right)}\exp\left(j\pi(\theta_{n'}^t-\theta_n^t)(N_t-1)\right),
\end{aligned}
\end{equation}
where $n$ and $n'$ indicate that two transmit beams are directed at two different scatterers.
Afterward, we take the absolute value of (\ref{prlem1}) and represent the operation as
\begin{equation}\label{prlem2}
\begin{aligned}
\left|\boldsymbol{a}^H_t(\theta_n^t)\boldsymbol{a}_t(\theta_{n'}^t)\right|
=&\frac{1}{N_t}\left|\frac{\sin\left(\pi(\theta_{n'}^t-\theta_n^t)N_t\right)}{\sin\left(\pi(\theta_{n'}^t-\theta_n^t)\right)}\right|.
\end{aligned}
\end{equation}
Without loss of generality, we consider that the scatterers involved in the modulation are all between the Tx and Rx channels, satisfying $(\theta_{n'}^t-\theta_n^t) \in \left[-\frac{\pi}{2},\frac{\pi}{2}\right]$.
Hence, we have
\begin{equation}
\left|\frac{\sin\left(\pi(\theta_{n'}^t-\theta_n^t)N_t\right)}{\sin\left(\pi(\theta_{n'}^t-\theta_n^t)\right)}\right|\neq 0.
\end{equation}
When $N_t \to \infty$, we have $\left|\boldsymbol{a}^H_t(\theta_n^t)\boldsymbol{a}_t(\theta_{n'}^t)\right|= 0$.
It is worth noting that we can obtain the corresponding proof for the receive array.
Here, the proof of {\bf Lemma 1} is completed.

\section{Proof of Lemma 2}%
It is clear to observe that the variables $g$ and $s$ are coupled together. To address this issue, we decompose $\eta$ into real part and the imaginary part.
After that, we discuss both cases $\hat{n}  ={n}$ and $\hat{n}  \neq{n}$ separately.

1) In {$\hat{n}  ={n}$} case,
we define $\eta = |g_ns_k-g_{\hat n}s_{\hat k}|^2$, then the average $\eta$ can be derived as
\begin{equation}\label{eta2}
\begin{aligned}
\bar{\eta}
=&E[{{|\zeta |^2}}]
+E[{{|\xi|^2}}]\\
=&E\left[\left|\Re\left(g_ns_k\right)
-\Re\left(g_{\hat n}s_{\hat k}\right)\right|^2\right]\\&
+E\left[\left|
\Im\left(g_ns_k\right)
-\Im\left(g_{\hat n}s_{\hat k}\right)\right|^2\right]\\=&E\left[\left|\Re\left(g_n\right)(s_k
-s_{\hat{k}})\right|^2\right]
+E\left[\left|
\Im\left(g_n\right)(s_k-s_{\hat{k}})\right|^2\right].
\end{aligned}
\end{equation}
For variable $\beta$, $x_\Re$ and $x_\Im$ can be viewed as constants.
Consequently, we have
\begin{equation}\label{eta4}
\begin{aligned}
\bar{\eta}=&E\left[\left|\Re\left(g_n\right)\right|^2\right]\left|s_k-s_{\hat{k}}\right|^2
+E\left[\left|
\Im\left(g_n\right)\right|^2\right]|s_k-s_{\hat{k}}|^2\\
=&E\left[
g_n^2\right]|s_k-s_{\hat{k}}|^2.
\end{aligned}
\end{equation}
Since $g_n$ follows $\mathcal{CN}(0,1)$, the $E\left[
g_n^2\right] = \mu_{g_n}^2+\sigma_{g_n}^2$.
At this time, we can obtain $\bar{\eta}
=|s_k-s_{\hat{k}}|^2$.

2) In {$\hat{n}  \neq{n}$} case,
the value of $\bar{\eta}$ can be we derived as
\begin{equation}\label{eta5}
\begin{aligned}
\bar{\eta}
=&E[{{|\zeta |^2}}]+E[{{|\xi|^2}}]\\
=&E\left[\left|\Re\left(g_n s_k\right)
-\Re\left(g_{\hat n} s_{\hat{k}}\right)\right|^2\right]\\&
+E\left[\left|
\Im\left(g_n s_k\right)
-\Im\left(g_{\hat n} s_{\hat{k}}\right)\right|^2\right]\\
=&E\left[\left|\Re\left(g_n\right)\right|^2\right]|s_{k}|^2
+E\left[\left|
\Re\left(g_{\hat n}\right)\right|^2\right]|{ s}_{\hat{k}}|^2\\&
+E\left[\left|\Im\left(g_n\right)\right|^2\right]|s_{k}|^2
+E\left[\left|
\Im\left(g_{\hat n}\right)\right|^2\right]|{x}_{\hat{k}}|^2\\
=&
\left(E\left[\left|\Re\left(g_n\right)\right|^2\right]
+E\left[\left|\Im\left(g_n\right)\right|^2\right]\right)|s_{k}|^2\\&
+\left(E\left[\left|
\Re\left(g_{\hat n}\right)\right|^2\right]
+E\left[\left|
\Im\left(g_{\hat n}\right)\right|^2\right]\right)|{s}_{\hat{k}}|^2\\
=&
E\left[g_n^2\right]|s_{k}|^2
+E\left[g_{\hat n} ^2\right]|{s}_{\hat{k}}|^2\\
=&
|s_{k}|^2
+|{s}_{\hat{k}}|^2.
\end{aligned}
\end{equation}
Herein, the proof of {\bf Lemma 2} is completed.

\section{Proof of Theorem 1}
Without loss of generality, let us define $x = |g_nh_m|^2$, where
$x_1 = |g_n|^2$ and $x_2 = |h_m|^2$.
Since both $x_1$ and $x_2$ follow independent and identically distributed $\mathcal{CN}(0,1)$, the cumulative distribution function (CDF) of $x$ can be derived as follows:
\begin{equation}\label{cdfX}
\begin{aligned}
F_X(x) &= P_r(X<x) = P_r(x_1x_2 \leq x)\\&
=\int_0^\infty P_r(x_1 \leq \frac{x}{x_2}|x_2)f(x_2)dx_2\\&=\int_0^\infty F\left(\frac{x}{x_2}\right)f(x_2)dx_2.
\end{aligned}
\end{equation}
Recall that $g_n \sim \mathcal{CN}(0,1)$ and $h_m \sim \mathcal{CN}(0,1)$.
After some mathematical operations, the {PDF and CDF} can be respectively given as
\begin{equation}\label{cdfX1}
F_{X_i}(x) = 1- \exp(-x), \ \ \ \ f_{X_i}(x) = \exp(-x), \ \ {i \in\{1,2\}}.
\end{equation}
By inserting (\ref{cdfX1}) into (\ref{cdfX}), we can further write (\ref{cdfX}) as
\begin{equation}\label{cdfX2}
\begin{aligned}
F_X(x)
&= 1-\int_0^\infty\exp\left(-\frac{x}{x_2}\right)\exp(-x_2)dx_2\\&=1-2\sqrt{x}K_1(2\sqrt{x}).
\end{aligned}
\end{equation}
Based on (\ref{cdfX2}), we can get
$
f(x) = \frac{d{F_X(x)}}{dx}.
$
At this time, the proof of {\bf Theorem 1} is completed.
\section{Proof of Theorem 2}%
By substituting
$
Q(x) = \frac{1}{\pi}\int_0^{\frac{\pi}{2}}\exp\left(-\frac{x^2}{2\sin^2\theta}\right)d\theta
$
into (\ref{pepe11}), the $A_1$ can be written as
\begin{equation}
A_1 = \frac{1}{\pi}\int_0^\infty\int_0^{\frac{\pi}{2}}
x^v \ln(x)\exp\left(-\frac{\rho L^2 \bar{\eta}x}{4\sin^2\theta}\right)d\theta dx.
\end{equation}
Exchange the order of integration of variables $\theta$ and $x$, $A_1$ can be further expressed as
\begin{equation}\label{pepa11}
A_1 = \frac{1}{\pi}\int_0^{\frac{\pi}{2}}\int_0^\infty
x^v \ln(x)\exp\left(-\frac{\rho L^2 \bar{\eta}x}{4\sin^2\theta}\right) dx d\theta.
\end{equation}
According to \cite{jef2007book}, we have
\begin{equation}\label{eq51}
\begin{aligned}
&\int_0^\infty x^v\ln(x)\exp\left(-\frac{ax}{b}\right)dx=\frac{b^{v+1}}{a^{v+1}}v!
\left[\psi(v+1)-\ln\left(\frac{a}{b}\right)\right].
\end{aligned}
\end{equation}
Substituting (\ref{eq51}) into (\ref{pepa11}), the $A_1$  can be represented as
\begin{equation}\label{a1e}
\begin{aligned}
A_1
=& \frac{1}{\pi}\int_0^{\frac{\pi}{2}}\int_0^\infty
x^v \ln(x)\exp\left(-\frac{\rho L^2 \bar{\eta}x}{4\sin^2\theta}\right) dx d\theta \\
=&\frac{1}{\pi}\int_0^{\frac{\pi}{2}}\left(\frac{4\sin^2\theta}{{\rho L^2 \bar{\eta}}}\right)^{v+1}v!
\left[\psi(v+1)-\ln\left(\frac{\rho L^2 \bar{\eta}}{4\sin^2\theta}\right)\right] d\theta \\
=&\frac{v!}{\pi}\left(\frac{4}{{\rho L^2 \bar{\eta}}}\right)^{v+1}\int_0^{\frac{\pi}{2}}\left({\sin\theta}\right)^{2v+2}\\&\times
\left[\psi(v+1)+\ln\left(\frac{4}{\rho L^2 \bar{\eta}}\right)+2\ln\left({\sin\theta}\right)\right] d\theta \\
=&\frac{v!}{\pi}\left(\frac{4}{{\rho L^2 \bar{\eta}}}\right)^{v+1}\left[\psi(v+1)+\ln\left(\frac{4}{\rho L^2 \bar{\eta}}\right)\right]\\&\times\underbrace{\int_0^{\frac{\pi}{2}}\left({\sin\theta}\right)^{2v+2}
 d\theta}_{B_1} \\
&+2\frac{v!}{\pi}\left(\frac{4}{{\rho L^2 \bar{\eta}}}\right)^{v+1}\underbrace{\int_0^{\frac{\pi}{2}}\left({\sin\theta}\right)^{2v+2}
\ln\left({\sin\theta}\right)d\theta}_{B_2}.
\end{aligned}
\end{equation}
In order to obtain the result of the integral, we refer to \cite{jef2007book} to deal with  $B_1$ and $B_2$ as
\begin{equation}\label{b1e}
B_1 = {\int_0^{\frac{\pi}{2}}(\sin\theta)^{2v+2}d\theta} = \frac{\pi}{2}\frac{(2v+1)!!}{(2v+2)!!},
\end{equation}
and
\begin{equation}\label{b1e2}
\begin{aligned}
B_2 &= \int_0^\infty \ln(\sin\theta)(\sin\theta)^{2v+2}d\theta\\&=\frac{\pi}{2}\frac{(2v+1)!!}{(2v+2)!!}\left[\sum_{q=1}^{2v+2}\frac{(-1)^{q+1}}{q}-\ln2\right].
\end{aligned}
\end{equation}
To facilitate the subsequent operations, we apply the following equivalent transformation as \cite{jef2007book}
\begin{equation}\label{sertgweg}
\sum_{q=1}^{2v+2}\frac{(-1)^{q+1}}{q} = \ln2 + \frac{1}{2}\left[
\psi\left(v+\frac{3}{2}\right)-\psi(v+2)
\right].
\end{equation}
Replacing (\ref{sertgweg}) with (\ref{b1e2}), we can rewrite $B_2$ as
\begin{equation}\label{B2e}
B_2 = \frac{\pi}{4}\frac{(2v+1)!!}{(2v+2)!!}\left[
\psi\left(v+\frac{3}{2}\right)-\psi(v+2)
\right].
\end{equation}
At this point, by substituting the derived (\ref{b1e}) and (\ref{B2e}) into the (\ref{a1e}), $A_1$ can be evaluated as
\begin{equation}\label{a1ee}
\begin{aligned}
A_1 &= \frac{v!}{2}\left(\frac{4}{{\rho L^2 \bar{\eta}}}\right)^{v+1}\frac{ (2v+1)!!}{(2v+2)!!}\\&\times\left[\psi(v+1)-\psi(v+2)+
\ln\left(\frac{4}{\rho L^2 \bar{\eta}}\right) + \psi\left(v+\frac{3}{2}\right)
\right].
\end{aligned}
\end{equation}
To simplify for (\ref{a1ee}), we obtain from \cite{jef2007book}
\begin{equation}\label{a1ee1}
\psi(v+1)-\psi(v+2) = -\frac{1}{v+1}.
\end{equation}
Finally, we insert (\ref{a1ee1}) into (\ref{a1ee}), the proof {\bf Theorem 2} is completed.

\section{Proof of Theorem 3}%
Substituting
$
Q(x) = \frac{1}{2}{\rm erfc}\left(\frac{x}{\sqrt{2}}\right)
$
into $A_2$, it can be rewritten as
\begin{equation}
\begin{aligned}
A_2
&=\frac{1}{2}\int_0^\infty x^v{\rm erfc}\left(\sqrt{\frac{\rho L^2 \bar{\eta}x}{4}}\right)
dx.
\end{aligned}
\end{equation}
Let us define $y = \sqrt{\frac{\rho L^2 \bar{\eta}x}{4}}$, $A_2$ can be rewritten as
\begin{equation}\label{sdfs0}
A_2 = {\left(\frac{4}{\rho L^2 \bar{\eta}}\right)^{v+1}}\int_0^\infty y^{2v+1}{\rm erfc}(y) dy.
\end{equation}
According to \cite{jef2007book}, we have
\begin{equation}
\int_0^\infty x^{2q-1}{\rm erfc}(x) dx = \frac{\Gamma(q+\frac{1}{2})}{2q\sqrt{\pi}}.
\end{equation}
Next, we substitute this term into (\ref{sdfs0}), the $A_2$ can be updated as
\begin{equation}\label{vv1}
A_2 = \frac{1}{2}\left(\frac{4}{\rho L^2 \bar{\eta}}\right)^{v+1}\frac{\Gamma(v+\frac{3}{2})}{(v+1)\sqrt{\pi}}.
\end{equation}
Resort to \cite{jef2007book}, we have
\begin{equation}\label{vv2}
\begin{aligned}
\Gamma\left(v+\frac{3}{2}\right) &= \frac{(2v+1)!!\sqrt{\pi}}{2^{v+1}},
\end{aligned}
\end{equation}
\begin{equation}\label{vv3}
\begin{aligned}
2^{v+1}=\frac{(2v+2)!!}{(v+1)!}.
\end{aligned}
\end{equation}
Substituting (\ref{vv2}) and (\ref{vv3}) into (\ref{vv1}), the proof of {\bf Theorem 3} is completed.

\section{Proof of (\ref{eq64}) and (\ref{eq65})}

To clarify the major parameters impacting the performance of ABEP, we resort to asymptotic expressions for our analysis. It is worth mentioning that we divide the received beams into two cases, correctly decoded and incorrectly decoded, for discussion.

{\bf Case 1}: If the detected beam is demodulated correctly, i.e., $\hat m = m$, the asymptotic UPEP of the SSM can be expressed as \cite{li2019polaried}
\begin{equation}\label{qssma}
\bar P_{\rm a}^{\rm SSM}  = \frac{24}{13\rho \bar{\eta}}.
\end{equation}
Recall that (\ref{asy1}), the asymptotic UPEP of RIS-DSSM scheme is given by
\begin{equation}\label{asy1e1}
\begin{aligned}
&\bar{P}_{\rm a}^{\rm RIS-DSSM}
 = \frac{1}{\rho L^2 \bar{\eta}}
\left[\ln\left(\frac{\rho L^2 \bar{\eta}}{4}\right) - \psi\left(\frac{3}{2}\right)+1
 +2\psi(1)\right].
\end{aligned}
\end{equation}
To obtain the minimal number of elements that RIS-DSSM better than SSM scheme, we let $\bar{P}_{\rm a}^{\rm RIS-DSSM} \leq \bar P_{\rm a}^{\rm SSM}$. After some manipulation operations, we have
\begin{equation}\label{eqle1}
\begin{aligned}
L^2- \frac{13}{ 24 }
\ln\left(\frac{\rho L^2 \bar{\eta}}{4}\right) +\frac{13}{ 24 }\left[\psi\left(\frac{3}{2}\right)-1
 -2\psi(1)\right]\leq0.
\end{aligned}
\end{equation}
By observation we find that the required variable $L$ contains $L^2$ and $\frac{13}{ 24 }
\ln\left(\frac{\rho L^2 \bar{\eta}}{4}\right)$ in two items, and it is impossible to directly obtain the exact value of $L$. However, through (\ref{eqle1}) we find that the factors affecting the RIS-DSSM and SSM schemes are not only related to the number of RIS cells but also to the SNR $\rho$.

{\bf Case 2}:
If the detected beam is demodulated incorrectly, i.e., $\hat m \neq m$, the CPEP of the SSM can be expressed as \cite{ding2017ssm}
\begin{equation}
P_b =\frac{1}{2}\exp\left(-\frac{\rho |h_m|^2|s_{\hat k}|^2}{2}\right).
\end{equation}
Based on this, we derive the UPEP of SSM as
\begin{equation}
\bar P_b^{\rm SSM} = \frac{1}{2}\int_0^\infty \exp\left(-\frac{\rho x|s_{\hat k}|^2}{2}\right)\exp(-x)dx =\frac{1}{\rho|s_{\hat k}|^2+2}.
\end{equation}
In the high SNR region, the constant term in the denominator can be ignored. At this point, we have
\begin{equation}
\bar P_a^{\rm SSM} =\lim\limits_{\rho\to \infty}\frac{1}{\rho|s_{\hat k}|^2+2}=\frac{1}{\rho|s_{\hat k}|^2}.
\end{equation}
Based on (\ref{asy2}), the asymptotic UPEP can be given as
\begin{equation}\label{asy2e1}
\begin{aligned}
&\bar{P}_{\rm a}^{\rm RIS-DSSM}
= \frac{1}{\rho L^2|s_{\hat k}|^2}\left[
\ln\left(\frac{\rho L^2|s_{\hat k}|^2}{2}\right)+{\psi(1)} \right].
\end{aligned}
\end{equation}
To obtain the minimal number of elements that RIS-DSSM better than SSM scheme, we let $\bar{P}_{\rm a}^{\rm RIS-DSSM} \leq \bar P_{\rm a}^{\rm SSM}$. Then, we have
\begin{equation}\label{asdfwe}
L^2-
\ln\left(\frac{\rho L^2|s_{\hat k}|^2}{2}\right)-{\psi(1)} \leq 0.
\end{equation}
Based on (\ref{asdfwe}), we find that the variable $L$ exists in $L^2$ and $\ln\left(\frac{\rho L^2|s_{\hat k}|^2}{2}\right)$ terms, which cannot be derived directly.
However, we observed that the impact of the RIS-DSSM and SSM schemes is related to both the number of RIS elements $L$ and $\rho$.

In summary, it is challenging to provide a precise quantitative assessment of the superiority of the RIS-DSSM system over the SSM system using analytical methods. Consequently, our paper primarily elucidates this distinction through simulation-based descriptions.

\end{appendices}

\end{document}